\documentclass{LMCS}

\usepackage[english]{babel}
\usepackage{graphics}
\usepackage{latexsym}
\usepackage{amsmath}
\usepackage{amsfonts}
\usepackage{amssymb}
\usepackage{enumerate}
\usepackage{epsf}
\usepackage{xypic}
\usepackage{hyperref}

\def\eqalign#1{\null\,\vcenter{\openup\jot\mathsurround=0 pt
  \ialign{\strut\hfil$\displaystyle{##}$&$\displaystyle{{}##}$\hfil
      \crcr#1\crcr}}\,}
\def\0#1 {{\mathstrut}_{\sf#1}}     \def\1#1 {{\mathstrut}^{\sf#1}}

\newcommand{\blast}{\textsc{Blast}}
\newcommand{\armc}{\textsc{ARMC}}
\newcommand{\ignore}[1]{}

\def\doi{4 (4:1) 2008}
\lmcsheading%
{\doi}
{1--31}
{}
{}
{Feb.~11, 2007}
{Oct.~17, 2008}
{}   

\begin{document}
\title{Interpolation in local theory extensions}
\author[V.~Sofronie-Stokkermans]{Viorica Sofronie-Stokkermans}
\address{Max-Planck-Institut f{\"u}r Informatik, Campus E1.4, 
Saarbr{\"u}cken, Germany}
\email{sofronie@mpi-inf.mpg.de}

\titlecomment{}
\begin{abstract}
In this paper we study interpolation in local extensions of a base theory. 
We identify situations in which it is possible 
to obtain interpolants in a hierarchical manner, 
by using a prover and a procedure for generating interpolants 
in the base theory as black-boxes. We present several examples 
of theory extensions in which interpolants can be computed this way, 
and discuss applications in verification, knowledge representation, 
and modular reasoning in combinations of local theories.
\end{abstract}
\maketitle

\section{Introduction}

\noindent Many problems in mathematics and computer science can be 
reduced to proving satisfiability of conjunctions of (ground) literals modulo 
a background theory. This theory can be a standard theory, 
the extension of a base 
theory with additional functions, or a combination of theories. 
It is therefore very important to find efficient 
methods for reasoning in standard as well as complex theories. 
However, it is often equally important to 
find  local  causes for inconsistency.
In distributed databases, for instance, finding local causes of 
inconsistency can help in locating errors. 
Similarly, in abstraction-based verification,
finding the cause of inconsistency in a counterexample at the concrete level 
helps to rule out certain spurious counterexamples in the abstraction.

\medskip

The problem we address in this paper can be described as follows: 
Let ${\mathcal T}$ be a theory and $A$ and $B$ be 
sets of ground clauses in the signature of ${\mathcal T}$, possibly 
with additional constants. 
Assume that $A \wedge B$ is inconsistent with respect to ${\mathcal T}$.
Can we find a ground formula $I$, 
containing only constants and function symbols common to 
$A$ and $B$, such that $I$ is a consequence 
of $A$ with respect to ${\mathcal T}$, and $B \wedge I$ is inconsistent modulo 
${\mathcal T}$?
If so, $I$ is a {\em (Craig) interpolant} of $A$ and $B$, and can be 
regarded as a ``local'' explanation for the inconsistency of $A \wedge B$.

\medskip

In this paper we study possibilities of obtaining ground interpolants in 
theory extensions. We identify situations in which it is possible to do 
this in a hierarchical 
manner, by using a prover and a procedure for 
generating interpolants in the base theory as ``black-boxes''. 

\medskip

We consider a special type of extensions of a base theory -- namely local 
theory extensions -- which 
we studied in \cite{Sofronie-cade-05}. We showed that in this case 
hierarchical reasoning is possible. i.e.\ proof tasks in the extension 
can be reduced to proof tasks in the base theory. Here we study 
possibilities of hierarchical interpolant generation in local 
theory extensions.

\medskip

The main contributions of the paper are summarized below: 
\begin{enumerate}[$\bullet$]
\item 
First, we identify new examples of local theory 
extensions. 
\item Second, we present a method for generating interpolants 
in local extensions of a base theory. 
The method is general, in the sense that it 
can be applied to an extension ${\mathcal T}_1$ of a theory ${\mathcal T}_0$ 
provided that: 
\begin{enumerate}[(i)]
\item ${\mathcal T}_0$ is convex;  
\item 
${\mathcal T}_0$ is $P$-interpolating for a specified set $P$ of predicates 
(cf.\ the definition in Section~\ref{examples-sep});  
\item in ${\mathcal T}_0$ every inconsistent conjunction of ground
  clauses $A \wedge B$ allows a ground interpolant;
\item the extension is defined by clauses of a special form 
(type (\ref{general-form}) in Section~\ref{examples-sep}).
\end{enumerate}

\smallskip
\noindent 
The method is {\em hierarchical}\/:  
the problem of finding interpolants 
in ${\mathcal T}_1$ is reduced to that of finding interpolants in the 
base theory ${\mathcal T}_0$. We can use the properties of 
${\mathcal T}_0$ to control the form of interpolants in the extension ${\mathcal T}_1$.

\smallskip
\item Third, we identify examples of theory extensions with 
properties (i)--(iv). 

\smallskip 
\item 
Fourth, we discuss application domains such as: 
modular reasoning in combinations of local theories (characterization of 
the type of information which needs to be exchanged),  
reasoning in distributed databases, and verification.
\end{enumerate}
The existence of ground interpolants has been studied in several 
recent papers, mainly motivated by abstraction-refinement based verification
\cite{McMillanCAV03,McMillanProver04,McMillanSurvey05,Yorsh-Musuvathi-cade-2005,Kapur-et-all-06}.
In \cite{McMillanProver04} McMillan 
presents a method for generating ground interpolants 
from proofs in an extension of linear rational arithmetic 
with uninterpreted function symbols. 
The use of free function symbols is sometimes too coarse
(cf.\ the example in Section~\ref{motivation-verif}). 
Here, we show that similar results also hold for other types of extensions 
of a base theory, provided that the base theory has some of the properties 
of linear rational arithmetic.  
Another method for generating interpolants for combinations of 
theories over disjoint signatures from Nelson-Oppen-style unsatisfiability 
proofs was proposed by Yorsh and Musuvathi in  
\cite{Yorsh-Musuvathi-cade-2005}. 
Although we impose similar conditions on 
${\mathcal T}_0$, our method is orthogonal to theirs, 
as it can also handle 
combinations of theories over non-disjoint signatures. 
In \cite{Kapur-et-all-06} a different interpolation property -- stronger 
than the property under consideration in this paper -- is 
studied, namely the existence of ground interpolants for 
{\em arbitrary formulae} -- which is proved to be equivalent to the theory 
having quantifier elimination. 
This limits the applicability of the results in \cite{Kapur-et-all-06} 
to situations in which the involved theories allow quantifier elimination. 
If the theory considered has quantifier elimination then we can 
use this for obtaining ground interpolants for arbitrary formulae. 
The goal of our paper is to identify theories -- possibly without 
quantifier elimination -- in which, nevertheless,  ground interpolants 
for ground formulae exist.

\medskip\noindent
{\em Structure of the paper:}\/  We start 
by providing motivation for the study in Section~\ref{motivation}. 
In Section~\ref{prelim} 
the basic notions needed in the paper are introduced. 
Section~\ref{local} contains results on local theory extensions. 
In Section~\ref{hierarchic} local extensions 
allowing hierarchical interpolation are identified, and based on this, 
in Section~\ref{procedure} a procedure for computing interpolants 
hierarchically is given.
In Section~\ref{appl} 
applications to modular reasoning in combinations of theories, reasoning in
complex databases, and verification are presented. 
In Section~\ref{conclusions} we draw conclusions, discuss the relationship 
with existing work, and sketch some plans for future work. 
For the sake of clarity in presentation, all the proofs that are not 
directly related to the main thread of the paper can be found 
in the appendix. (These results concern illustrations 
of the fact that certain 
theory extensions are local, or satisfy assumptions that 
guarantee that interpolants can be computed hierarchically.)

\section{Motivation}
\label{motivation}

\noindent 
In this section we present two fields of applications in which it is important to efficiently compute interpolants: knowledge representation 
and verification.

\subsection{Knowledge representation} 
\label{motivation-knowledge}
Consider a simple (and faulty) terminological database for chemistry,
consisting of two extensions of a common kernel {\sf Chem} (basic
chemistry): {\sf AChem} (inorganic (anorganic) chemistry) and {\sf
  BioChem} (biochemistry).  Assume that {\sf Chem} contains a set
${\sf C}_0$ $=$ $\{ {\sf process},$ ${\sf reaction}, {\sf substance},$
${\sf organic}, $ ${\sf inorganic} \}$ of concepts and a set
$\Gamma_0$ of constraints:
\[ 
{\Gamma}_0 = \{ {\sf organic} \wedge {\sf inorganic} = \emptyset,
\enspace {\sf organic} \subseteq {\sf substance}, \enspace {\sf inorganic}
\subseteq {\sf substance} \}\ . 
\] 
Let {\sf AChem} be an extension of {\sf Chem} with concepts ${\sf C}_1
= \{ {\sf cat}$-${\sf oxydation}, {\sf oxydation} \}$, a r{\^o}le
${\sf R}_1 = \{ {\sf catalyzes} \}$, terminology ${\sf T}_1$ and
constraints $\Gamma_1$:
\[\eqalign{
  {\sf T}_1
&=\{ {\sf cat\hbox{-}oxydation} = {\sf substance} 
    \wedge {\exists}\,{\sf catalyzes}({\sf oxydation})\}\cr
  \Gamma_1 
&=\{ {\sf reaction} \subseteq {\sf oxydation}, \enspace 
     {\sf cat\hbox{-}oxydation} \subseteq {\sf inorganic}, \enspace 
     {\sf cat\hbox{-}oxydation} \neq \emptyset \}\ .\cr
}
\]
Let {\sf BioChem} be an extension of {\sf Chem} with the  
concept ${\sf C}_2 = \{ {\sf enzyme} \}$, the r{\^o}les 
${\sf R}_2 = \{ {\sf produces}, {\sf catalyzes} \}$, 
terminology ${\sf T}_2$ and constraints $\Gamma_2$:
\[\eqalign{
  {\sf T}_2 
&=\{ {\sf reaction} {=} {\sf process} \wedge 
{\exists}\,{\sf produces}({\sf substance}), {\sf enzyme} = {\sf organic} \wedge 
{\exists}\,{\sf catalyzes}({\sf reaction}) \} \cr
  \Gamma_2 
&=\{ {\sf enzyme} \neq \emptyset \}\ .
}
\] 
The combination  of {\sf Chem}, {\sf AChem} and {\sf BioChem} 
is inconsistent (we wrongly added to $\Gamma_1$ the constraint 
${\sf reaction} \subseteq {\sf oxydation}$ instead of ${\sf oxydation} \subseteq {\sf reaction}$).
This can be proved as follows: 
By results in \cite{Sofronie-jim-03} (p.156 and p.166) the combination 
 of {\sf Chem}, {\sf AChem} and {\sf BioChem} is inconsistent if and only if  
\begin{eqnarray}
\Gamma_0 \wedge ({\sf T}_1  \wedge {\Gamma_1}) \wedge ({\sf T}_2 \wedge {\Gamma_2}) \models_{\mathcal T} \perp \label{incons-s1}
\end{eqnarray}
where ${\mathcal T}$ is the extension 
${\sf SLat} \wedge \bigcup_{f \in {\sf R}_1 \cup {\sf R}_2} {\sf Mon}(f)$ of the theory of 
semilattices with smallest element 
0 and 
monotone function symbols corresponding to $\exists r$ for each r{\^o}le  
$r \in {\sf R}_1 \cup {\sf R_2}$. 
Using, for instance, the hierarchical calculus presented in 
\cite{Sofronie-cade-05} (see also Section~\ref{local}), 
the contradiction can be found in polynomial time. 
In order to find the mistake 
we look for an explanation for the inconsistency in the common language
of ${\sf AChem}$ and ${\sf BioChem}$. 
(Common to ${\sf AChem}$ and ${\sf BioChem}$ 
are the concepts ${\sf substance}, {\sf organic}, 
{\sf inorganic}, {\sf reaction}$ and the r{\^o}le ${\sf catalyzes}$.) 
This can be found by computing an interpolant for 
the conjunction in~(\ref{incons-s1}) 
in the theory of semilattices with monotone operators.
In this paper we show how such interpolants can be found in an efficient way.
The method is illustrated on the example above in Section~\ref{appl-knowledge}.

\subsection{Verification}
\label{motivation-verif}
In \cite{McMillanProver04}, McMillan  proposed a method for abstraction-based 
verification in which 
interpolation (e.g.\ for linear arithmetic + free functions) 
is used for abstraction refinement. The idea is the following: 
Starting from a concrete, precise description of a 
(possibly infinite-state) system one can obtain a finite abstraction, 
by merging the states into equivalence classes. A transition exists 
between two abstract states if there exists a transition in the 
concrete systems between representatives in the corresponding equivalence
classes. Literals describing 
the relationships between the state variables at the concrete level are 
represented -- at the abstract level -- by predicates on the abstract 
states (equivalence classes of concrete states). 
Classical methods (e.g. BDD-based methods) can be used for checking 
whether there is a path in the abstract model from an 
initial state to an unsafe state. We distinguish the following cases:
\begin{enumerate}[(1)]
\item No unsafe state is reachable from 
an initial state in the abstract model. Then, 
due to the way transitions are defined 
in the abstraction, this is the case also at the concrete level. 
Hence, the concrete system is guaranteed to be safe.
\item There exists a path in the abstract model from an 
initial state to an unsafe state. 
This path may or may not have a correspondent at the concrete level. 
In order to check this, we analyse the counterpart of the counterexample
in the concrete model. This can be reduced to testing the 
satisfiability of a set of constraints: 
\[ {\sf Init}(s_0) \wedge {\sf Tr}(s_0, s_1) \wedge \dots \wedge {\sf Tr}(s_{n-1}, s_{n}) \wedge \neg {\sf Safe}(s_n) \]
\begin{enumerate}[(2.1)]
\item
If the set of constraints is satisfiable then an unsafe state is reached from 
the initial state also in the concrete system.
Thus, the concrete system is not safe. 
\item
If the set of constraints is unsatisfiable, then the counterexample 
obtained due to the abstraction was spurious. This means that the abstraction
was too coarse. In order to refine it we need to take into account 
new predicates or relationships between the existing predicates.
Interpolants provide information about which new predicates need 
to be used for refining the abstraction.
\end{enumerate}
\end{enumerate}

\smallskip
\noindent We illustrate these ideas below. Consider a water 
level controller modeled as follows: Changes in the water 
level by inflow/outflow are represented as functions 
${\sf in}, {\sf out}$, depending on
time $t$ and water level $L$. Alarm and overflow levels $L_{\sf alarm} {<} 
L_{\sf overflow}$, as well as upper/lower bounds for mode durations 
$0 \leq \delta t \leq \Delta t$  are parameters of the systems.
\bigskip\bigskip

{\noindent
\leavevmode
\epsfverbosetrue %
\def\epsfsize#1#2{0.4#1}              
\begin{tabular}{ll}
\epsffile{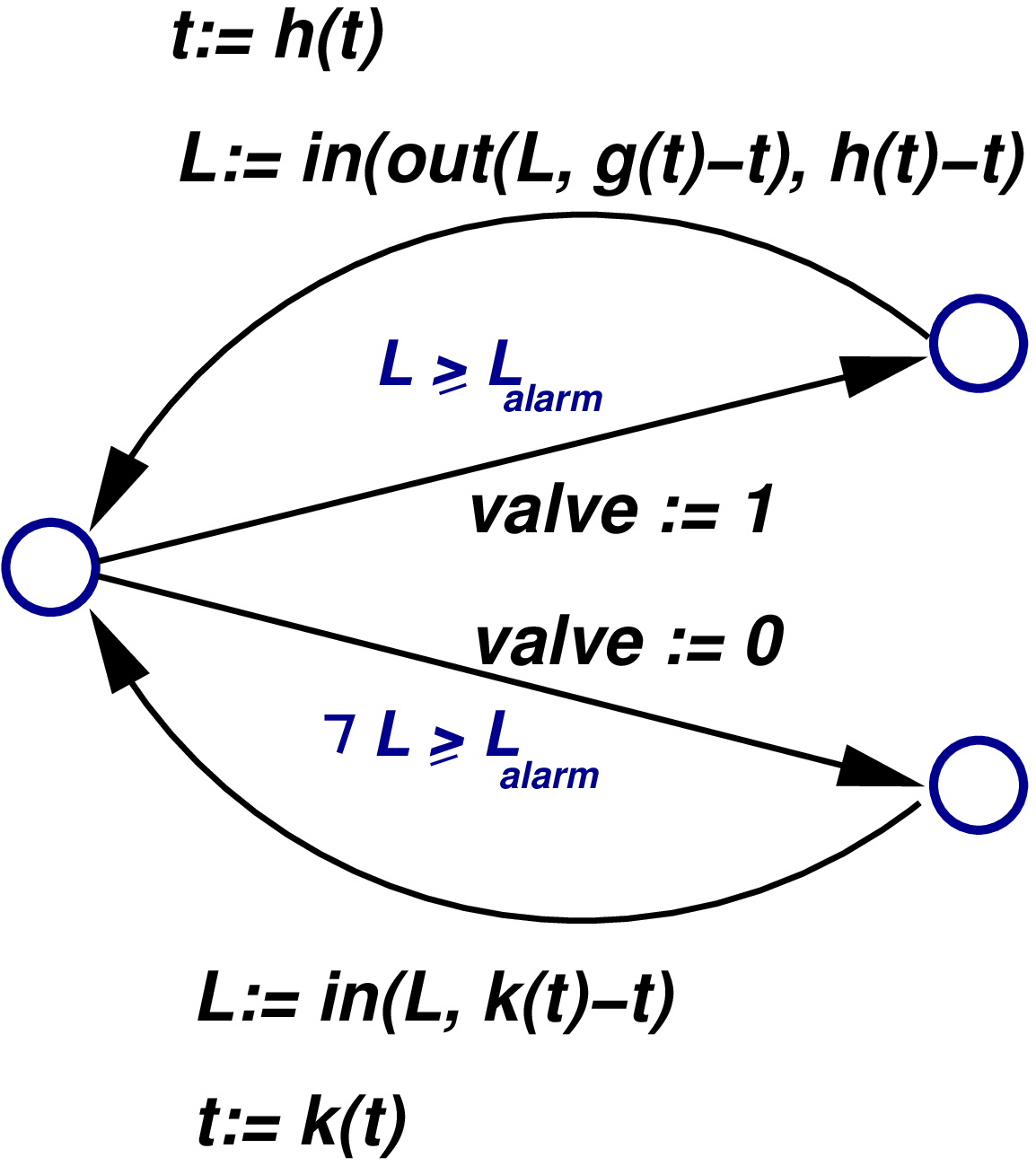} 
& 
\begin{tabular}{l}

\vspace{-4.8cm}
\hspace{1.8cm}\\

$\bullet$ If $L \geq L_{\sf alarm}$ then a valve is opened until time $g(t)$,\\
~~ time changes to $t' := h(t)$ and the water level to \\
~~ $L' := {\sf in}({\sf out}(L, g(t) - t), h(t) - t)$.\\[1.2ex]
$\bullet$  If $L < L_{\sf alarm}$  then the valve is closed; time changes to \\
~~~ $t' := k(t)$, and the water level to $L' := {\sf in}(L, k(t) - t)$.\\[2ex]

We impose restrictions ${\mathcal K}$ on $h, g, k$ and on ${\sf in}$ and 
${\sf out}$:\\[1ex]

$~~~\forall t~~ (0 \leq \delta t \leq g(t) - t \leq h(t) - t \leq \Delta t)$\\
$~~~\forall t~~ (0 \leq k(t) - t \leq \Delta t) $\\ 
$~~~\forall L, t~ (L < L_{\sf alarm} \wedge 0 \leq t \leq \Delta t \rightarrow  {\sf in}(L, t) < L_{\sf overflow}) $ \\ 
$~~~\forall L, t~ (L < L_{\sf overflow} \wedge t \geq \delta t \rightarrow   {\sf out}(L, t) < L_{\sf alarm}). 
\label{1}$ \\[2ex]

We want to show that if initially $L < L_{\sf alarm}$ then the \\
water level always remains below $L_{\sf overflow}$.\\[1ex]
 \end{tabular} 
\end{tabular}
}

\medskip
\noindent 
We start with an abstraction in which the predicates are: 
$$\begin{array}{llllll}
p: & L < L_{\sf alarm} & & & &  \\      
r_1: & t' \approx k(t) & r_2: & t''_1 \approx g(t')  
& r_3: & t''_2 \approx h(t')   \\
p_1: & L' \approx {\sf in}(L, t' - t) & p_2: & L' \geq L_{\sf alarm} 
& p_3: & L'' \approx {\sf in}({\sf out}(L', t''_1 - t'), t''_2 - t')   \\
q: & \neg L'' < L_{\sf overflow}  & & & & \\
\end{array}$$ 
and no other relations between these predicates are specified.
We can, for instance, use finite model checking for the finite abstraction 
obtained this way. Note for instance that 
$$p \wedge p_1 \wedge p_2 \wedge p_3 \wedge r_1 \wedge r_2 \wedge r_3 
\wedge q$$
is satisfiable, i.e. in the abstract model there exists 
a path (of length 2) from the initial state to an unsafe state. 
We analyze the corresponding path in the concrete model to see if this 
counterexample to safety is spurious, i.e. we check 
whether there exist 
$l, l', l'', t, t', t''_1, t''_2 \in {\mathbb R}$ 
such that the conjunction:
\begin{eqnarray*} 
G & = &  l {<} L_{\sf alarm} ~\wedge~ l' \approx {\sf in}(l, t' - t) ~\wedge~ t' \approx k(t) ~\wedge~ l' \geq L_{\sf alarm} ~\wedge~ \\
& & l'' \approx {\sf in}({\sf out}(l', t''_1 - t'), t''_2 - t') ~\wedge~ t''_1 \approx g(t') ~\wedge~ t''_2 \approx h(t')
~\wedge~  \neg l'' < L_{\sf overflow}
\end{eqnarray*}
is true. 
If $h, g, k, {\sf in}, {\sf out}$ are regarded as free function symbols
this conjunction is satisfiable, so the spuriousness of the counterexample
cannot be detected. $G$ can however be proved to be 
unsatisfiable if we take into account the additional conditions ${\mathcal K}$ 
on the functions 
${\sf in}, {\sf out}, g, h$ and $k$. Interpolants can be used for 
determining the cause of inconsistency, and can therefore help in 
refining the abstraction. 
The hierarchical interpolation method we present here allows us to 
efficiently generate ground interpolants for extensions with functions 
satisfying axioms of the type considered here and also for a whole 
class of more general axioms. 
An illustration of this method on the formulae in 
the example presented here is given in Section~\ref{appl-verification}.

\smallskip
Besides the application to verification by abstraction-refinement,  
computation of Craig interpolants has other potential applications (e.g.\ 
to goal-directed overapproximation for achieving faster termination, or 
to automatic invariant generation).

\section{Preliminaries}
\label{prelim}

\noindent In this section we introduce the 
main notions and definitions concerning theories, models and interpolants 
needed in the paper. 
\subsection{Theories and models}
Theories can be regarded as sets of formulae or as sets of models. 
In this paper, whenever we speak about a theory ${\mathcal T}$  
-- if not otherwise 
specified -- we implicitly refer to the set 
${\sf Mod}({\mathcal T})$ of all models of ${\mathcal T}$. 
\begin{defi}
Let ${\mathcal T}$ be a theory in a given signature $\Pi = (\Sigma, {\sf Pred})$, 
where $\Sigma$ is a set of function symbols and ${\sf Pred}$ a set of 
predicate symbols. 
Let $\phi$ and $\psi$ be formulae over the signature 
$\Pi$ with variables in a set $X$. 
The notion of truth of formulae and of entailment is the usual one in 
logic. We say that:
\begin{enumerate}[$\bullet$]
\item $\phi$ is true with respect to ${\mathcal T}$ 
(denoted $\models_{\mathcal T} \phi$) if $\phi$ is true in each model 
${\mathcal M}$  of ${\mathcal T}$. 
\item $\phi$ is satisfiable with respect to ${\mathcal T}$  if there exists at 
least one model 
${\mathcal M}$ of ${\mathcal T}$ and an assignment 
$\beta : X \rightarrow {\mathcal M}$ 
such that $({\mathcal M}, \beta) \models \phi$. Otherwise 
we say that $\phi$ is unsatisfiable. 
\item We say that $\phi$ entails $\psi$ with respect to ${\mathcal T}$ 
(denoted $\phi \models_{\mathcal T} \psi$) 
if for every model ${\mathcal M}$ of ${\mathcal T}$ and every valuation 
$\beta$, 
if $({\mathcal M}, \beta) \models \phi$ then 
$({\mathcal M}, \beta) \models \psi$. 
\end{enumerate}
\end{defi}
\noindent Note that 
$\phi$ is unsatisfiable with respect to ${\mathcal T}$ if and only if  
$\phi \models_{\mathcal T} \perp$ ~($\perp$ stands for {\sf false}).

\subsection{Interpolation}
A theory ${\mathcal T}$ has interpolation if, for all
formulae  $\phi$ and $\psi$ in the signature of ${\mathcal T}$, 
if $\phi \models_{\mathcal T} \psi$ then there exists a formula 
$I$ containing only symbols which occur in both $\phi$ and 
$\psi$ such that  $\phi \models_{\mathcal T} I$ and 
$I  \models_{\mathcal T} \psi$. 
First order logic has interpolation but -- for an arbitrary theory 
${\mathcal T}$ -- even if 
$\phi$ and $\psi$ are e.g. conjunctions of ground literals,    
$I$ may still be an arbitrary formula, containing alternations of quantifiers
(cf.\ \cite{Kapur-et-all-06} for an example of 
ground formulae $\phi$ and $\psi$ in the language of 
the theory ${\sf Th}_{\sf arrays}$ of arrays 
whose conjunction is unsatisfiable, but there 
is no ground interpolant over the common variables of $\phi$ and $\psi$).

\smallskip
\noindent It is often important to identify situations in which  
ground clauses have ground interpolants. 

\begin{defi}
We say that a theory ${\mathcal T}$ has the {\em ground interpolation 
property} (or, shorter, that ${\mathcal T}$ has {\em ground interpolation}) 
if for all ground 
clauses $A({\overline c}, {\overline d})$ and  
$B({\overline c}, {\overline e})$,  if 
$A({\overline c}, {\overline d}) \wedge B({\overline c}, {\overline e}) \models_{\mathcal T} \perp$ then 
there exists a ground formula $I({\overline c})$, containing only 
the constants ${\overline c}$ occurring both in $A$ and $B$, such that 
$A({\overline c}, {\overline d}) \models_{\mathcal T} I({\overline c}) 
\text{ and } 
B({\overline c}, {\overline e}) \wedge 
I({\overline c}) \models_{\mathcal T} \perp.$
\end{defi} 

There exist results which relate ground interpolation to 
amalgamation or the injection transfer property 
\cite{Jonsson65,Bacsich75,Wronski86} 
and thus allow us to recognize 
many theories with ground interpolation.  We present these results in 
Appendix~\ref{app:amalg-interp}. 

\begin{thm} 
The following theories allow ground interpolation\footnote{In fact, 
the theories (1) and (4) allow  equational interpolation 
(cf.\ Definition~\ref{equational-interpolation} in
 Appendix~\ref{app:amalg-interp}). Similar results were also 
established for (2)  in \cite{RybalchenkoSofronie06}.}: 
\begin{enumerate}[\em(1)]
\item The theory of pure equality (without function symbols). 
\item Linear rational and real arithmetic.
\item The theory of posets.
\item The theories of (a) Boolean algebras, (b) semilattices, (c) 
distributive lattices.
\end{enumerate}
\label{ground-interp-eq-th}
\end{thm} 
\Proof The proof is given in Appendix~\ref{app:amalg-interp}. \qed

\smallskip
Other examples of theories which allow ground interpolation are 
the equational classes of 
(abelian) groups and lattices.
In many applications one needs to consider extensions or 
combinations of theories, and proving amalgamation properties 
can be complicated. On the other hand, just knowing that ground 
interpolants exist is usually not sufficient: we would like to 
construct the interpolants fast. 

\smallskip
In the examples considered in Theorem~\ref{ground-interp-eq-th}, methods 
for constructing interpolants exist. For the theories of pure equality and 
of posets interpolants can be constructed for instance from proofs 
\cite{McMillanProver04,Yorsh-Musuvathi-cade-2005}. For linear rational or real 
arithmetic they can either be constructed from proofs \cite{McMillanProver04}
or by constructing linear programming problems and solving these problems 
using an off-the-shelf sound solver \cite{RybalchenkoSofronie06}
\footnote{Some off-the-shelf linear programming solvers may 
not be sound, so care is needed when choosing them.}. 
For the theories of Boolean algebras, distributive lattices and semilattices
they can be reconstructed from resolution proofs associated with the 
translation of the satisfiability problems to propositional logic 
\cite{Sofronie-jsc-2003}; the construction is similar to the one 
described in the proof of Theorem~\ref{example-assumptions-1-3} in 
Appendix~\ref{app:assumptions-examples}. 

\smallskip
\noindent 
We would like to use the advantages of modular or hierarchical reasoning for 
constructing interpolants in theory extensions in an efficient way. 
This is why in this paper we aim at 
giving methods for {\em constructing} interpolants in a hierarchical way.
Since in \cite{Sofronie-cade-05} we identified a class of theory 
extensions -- namely, local theory extensions -- 
in which hierarchical reasoning was possible, in what follows we will 
study interpolation in local theory extensions.

\section{Local Theory Extensions}
\label{local}

\noindent Let ${\mathcal T}_0$ be a theory with signature 
$\Pi_0 = (\Sigma_0, {\sf Pred})$. 
We consider extensions
${\mathcal T}_1 = {\mathcal T}_0 \cup {\mathcal K}$ of 
${\mathcal T}_0$  with signature 
$\Pi = ({\Sigma}, {\sf Pred})$, 
where $\Sigma = \Sigma_0 \cup \Sigma_1$ 
(i.e.\ the 
signature is extended by new {\em function symbols}) and 
${\mathcal T}_1$ is obtained from ${\mathcal T}_0$ by 
adding a set ${\mathcal K}$ of (universally quantified) clauses.
Thus, ${\sf Mod}({\mathcal T}_1)$ consists of all $\Pi$-structures 
${\mathcal M}$ which are models of ${\mathcal K}$ and 
whose reduct ${\mathcal M}_{|\Pi_0}$ to $\Pi_0$ is a model of ${\mathcal T}_0$.

\begin{defi}
A {\em partial $\Pi$-structure} is a structure 
${\mathcal M} = (M, \{ f_M \}_{f \in \Sigma}, \{ P_M \}_{P \in {\sf Pred}})$, 
where $M \neq \emptyset$ 
and for every $f \in \Sigma$ with arity $n$, 
$f_M$ is a partial function from $M^n$ to $M$. 

\end{defi}
Any variable assignment $\beta :  X \rightarrow M$ extends in a natural 
way to terms, such that 
$\beta(f(t_1, \dots, t_n)) = f_M(\beta(t_1), \dots, \beta(t_n))$. 
Thus, the notion of evaluating a term $t$ with respect to a variable 
assignment 
$\beta : X \rightarrow M$ for its variables in a partial structure 
${\mathcal M}$ 
is the same as for total algebras, except that this evaluation is undefined
if $t = f(t_1, \dots, t_n)$ and at least one of $\beta(t_i)$ is undefined, 
or else $(\beta(t_1), \dots, \beta(t_n))$ is not in the domain of $f_M$.  
\begin{defi}
Let ${\mathcal M}$ be a partial $\Pi$-structure, $C$ a clause and 
$\beta : X \rightarrow M$. 
Then $({\mathcal M}, \beta) \models_w C$ if and only if  either 
\begin{enumerate}[(i)]
\item
for some term $t$ in $C$,
$\beta(t)$ is undefined, or else 
\item $\beta(t)$ 
is defined for all terms $t$ of $C$, 
and there exists a literal $L$ in $C$ such that $\beta(L)$ is true 
in ${\mathcal M}$.
\end{enumerate} 
{\em ${\mathcal M}$ 
weakly satisfies $C$} (notation: ${\mathcal M} \models_w C$) 
if $({\mathcal M}, \beta) \models_w C$
for all assignments $\beta$.  
We say that 
{\em ${\mathcal M}$ weakly satisfies a set of clauses ${\mathcal K}$} 
or {\em ${\mathcal M}$ is a weak partial model of ${\mathcal K}$} 
(notation: ${\mathcal M} \models_w {\mathcal K}$) if 
${\mathcal M} \models_w C$ for all $C \in {\mathcal K}$.
\end{defi}
\subsection{Local theory extensions: definitions}
Let ${\mathcal T}_0$ be a theory with signature 
$\Pi_0 = (\Sigma_0, {\sf Pred})$ 
and let ${\mathcal K}$ be a set of (universally
quantified)  clauses in the signature 
$\Pi = (\Sigma, {\sf Pred})$, where $\Sigma = \Sigma_0 {\cup} \Sigma_1$.
In what follows, when referring to sets $G$ of ground clauses 
we assume they are in the signature 
$\Pi^c = (\Sigma {\cup} \Sigma_c, {\sf Pred})$
where $\Sigma_c$ is a set of new constants.  
For the sake of simplicity, we will use the same 
notation for a structure and for its universe. 

\smallskip
A (total) model of ${\mathcal  T}_1 = {\mathcal T}_0 \cup {\mathcal K}$ 
is a $\Pi$-structure $A$ s.t.\
$A \models {\mathcal  K}$ and $A_{|\Pi_0}$ is a model of ${\mathcal  T}_0$. 
Let ${\sf PMod_w}({\Sigma_1}, {\mathcal  T}_1)$ 
be the class of all weak 
partial models $P$ of ${\mathcal  K}$, in which the 
$\Sigma_1$-functions are partial and 
such that $P_{|\Pi_0}$ is a total model of ${\mathcal  T}_0$. 

\smallskip
An extension ${\mathcal T}_0 \subseteq {\mathcal T}_0 \cup {\mathcal K}$ 
is {\em local} if, in order to prove unsatisfiability of a set 
$G$ of clauses with respect to ${\mathcal T}_0 \cup {\mathcal K}$, 
it is sufficient to use only those instances ${\mathcal K}[G]$ of 
${\mathcal K}$ in which the terms starting with extension functions are 
in the set ${\sf st}(G, {\mathcal K})$ of ground terms which already  occur
in $G$ or ${\mathcal K}$. 

\begin{defi}
We consider the following properties of an extension 
${\mathcal T}_1 {=} {\mathcal T}_0 \cup {\mathcal K}$ of a theory 
${\mathcal T}_0$ with additional function symbols satisfying 
a set ${\mathcal K}$ of clauses. 

\smallskip
\noindent
\begin{tabular}{ll}
${\sf (Loc)}\!\!$  & For every set $G$ of ground clauses,  
$G \models_{{\mathcal T}_1}  \perp$ if and only if  there is no partial \\
& $\Pi^c$-structure $P$ such that $P_{|\Pi_0}$ is a total 
model of ${\mathcal T}_0$, all terms in ${\sf st}({\mathcal K}, G)$ are \\
& defined in $P$, 
and $P$ weakly satisfies ${\mathcal K}[G] \wedge G$. \\[1ex]
\end{tabular}

\smallskip
\noindent A weaker notion ${\sf (Loc^f)}$ is defined by requiring that the 
locality condition only holds for {\em finite} sets $G$ of ground clauses.

\smallskip
\noindent
\begin{tabular}{ll}
${\sf (Loc^f)}\!\!$  & For every {\em finite} set $G$ of ground clauses,  
$G \models_{{\mathcal T}_1}  \perp$ if and only if  there is no partial \\
& $\Pi^c$-structure $P$ such that $P_{|\Pi_0}$ is a total 
model of ${\mathcal T}_0$, all terms in ${\sf st}({\mathcal K}, G)$ are \\
& defined in $P$, 
and $P$ weakly satisfies ${\mathcal K}[G] \wedge G$. \\[1ex]
\end{tabular}
\end{defi}

\smallskip
\noindent Since ${\sf (Loc^f)}$ is the property we are interested in, we will only 
refer to this form of locality in what follows. We will say that the 
extension ${\mathcal T}_0 \subseteq {\mathcal T}_1$ is {\em local} if it satisfies 
condition ${\sf (Loc^f)}$.

\subsection{Embeddability and locality.} 
In \cite{Sofronie-cade-05,Sofronie-Ihlemann07} we showed that 
embeddability of certain weak partial models into 
total models implies locality of an extension. 
Consider condition: 

\smallskip
\noindent \begin{tabular}{ll}
${\sf (Emb^{fd}_w)}$  & Every $A \in {\sf PMod_w}({\Sigma_1}, {\mathcal T}_1)$ 
in which all extension functions have a \\
& {\em finite definition domain} weakly embeds into a total model of ${\mathcal T}_1$. 
\end{tabular}

\begin{defi}
A non-ground clause is $\Sigma_1$-{\em flat} 
if function symbols (including constants) do not occur 
as arguments of functions in $\Sigma_1$.
A $\Sigma_1$-flat non-ground clause is called $\Sigma_1$-{\em linear} 
if whenever a variable occurs in two terms in the clause 
which start with function symbols in $\Sigma_1$, 
the two terms are identical, and if 
no term which starts with a function 
in $\Sigma_1$ contains two occurrences of the same variable.
\end{defi}
\begin{thm}[\cite{Sofronie-cade-05,Sofronie-Ihlemann07}]
Let ${\mathcal K}$ be a set of clauses in which all terms starting with a 
function symbol in $\Sigma_1$ are flat and linear. 
If the extension ${\mathcal T}_0 \subseteq {\mathcal T}_1$ satisfies 
${\sf (Emb^{fd}_w)}$ then it satisfies ${\sf (Loc^f)}$.
\label{rel-loc-embedding}
\end{thm}

\subsection{Examples}
\label{sect:local-examples}
Using a variant of Theorem~\ref{rel-loc-embedding}, 
in \cite{Sofronie-cade-05} we gave several examples of local 
theory extensions: 
any extension of a theory with free functions; 
extensions with selector functions for a constructor which 
is injective in the base theory; extensions of ${\mathbb R}$ with a 
Lipschitz function in a point $x_0$; extensions of partially ordered 
theories -- in a class ${\sf Ord}$ 
consisting of the theories of posets, 
(dense) totally-ordered sets, semilattices, 
(distributive) lattices, Boolean algebras, or 
${\mathbb R}$ --  with 
a monotone function $f$, i.e.\ satisfying: 

\smallskip
~~~~~~~~~~~~~$ ({\sf Mon}(f)) \quad \quad \displaystyle{\bigwedge_{i = 1}^n} x_i \leq y_i \rightarrow f(x_1, \dots, x_n) \leq f(y_1, \dots, y_n).$

\smallskip
\noindent 
Generalized monotonicity conditions -- combinations of 
monotonicity in some arguments 
and antitonicity in other arguments  -- were studied in 
\cite{Sofronie-Ihlemann07}. 
Below, we give some additional examples with 
particular relevance in verification. 
\begin{thm}
We consider the following base theories ${\mathcal T}_0$: 
(1) ${\mathcal P}$ (posets), 
(2) ${\mathcal TO}$ (totally-ordered sets), 
(3) ${\sf SLat}$ (semilattices), 
(4) ${\sf DLat}$ (distributive lattices), 
(5) ${\sf Bool}$ (Boolean algebras),  
(6) the theory ${\mathbb R}$ of reals resp.\ 
${\sf LI}(\mathbb R)$ (linear arithmetic over ${\mathbb R}$), or 
the theory ${\mathbb Q}$ of rationals resp.\  
${\sf LI}(\mathbb Q)$ (linear arithmetic over ${\mathbb Q}$),   
or (a subtheory of) the theory of integers (e.g.\ Presburger arithmetic). 
The following theory extensions are local:
\begin{enumerate}[\em(a)] 
\item Extensions of any theory ${\mathcal T}_0$ 
for which $\leq$ is reflexive with functions satisfying boundedness 
$({\sf Bound}^t(f))$ or guarded boundedness $({\sf GBound}^t(f))$ conditions 

\smallskip
$({\sf Bound}^t(f)) \quad  \quad \forall x_1, \dots, x_n (f(x_1, \dots, x_n) \leq t(x_1, \dots, x_n))$ 

$({\sf GBound}^t(f)) \quad \forall x_1, \dots, x_n (\phi(x_1, \dots, x_n) \rightarrow f(x_1, \dots, x_n) \leq t(x_1, \dots, x_n)),$

\smallskip
\noindent 
where $t(x_1, \dots, x_n)$ is a term in the base signature $\Pi_0$ and 
$\phi(x_1, \dots, x_n)$ a conjunction of literals in the signature $\Pi_0$, 
whose variables are in $\{ x_1, \dots, x_n \}$.

\smallskip 
\item Extensions of any theory  ${\mathcal T}_0$  in (1)--(6)  
with ${\sf Mon}(f) \wedge {\sf Bound}^t(f)$, if $t(x_1, \dots, x_n)$ is 
a term in the base signature $\Pi_0$ in the variables $x_1, \dots, x_n$ 
such that for every model of ${\mathcal T}_0$ the associated function is
monotone in the variables $x_1, \dots, x_n$.

\smallskip
\item Extensions of any theory in  (1)--(6) 
with functions satisfying ${\sf Leq}(f,g) \wedge {\sf Mon}(f)$.

\smallskip
$({\sf Leq}(f,g)) \quad \forall x_1, \dots, x_n (\bigwedge_{i = 1}^n x_i \leq y_i \rightarrow f(x_1, \dots, x_n) \leq g(y_1, \dots, y_n))$

\smallskip
\item Extensions of any totally-ordered theory above (i.e.\ (2) and (6))
with functions satisfying 
${\sf SGc}(f,g_1, \dots, g_n) \wedge {\sf Mon}(f, g_1, \dots, g_n)$.

\smallskip
$({\sf SGc}(f,g_1, \dots, g_n)) \quad \forall x_1,\dots, x_n, x ( \bigwedge_{i = 1}^n x_i  \leq g_i(x) \rightarrow f(x_1, \dots, x_n) \leq x)$

\smallskip
\item Extensions of  any theory in (1)--(3) 
with functions satisfying ${\sf SGc}(f,g_1) \wedge {\sf Mon}(f, g_1)$.
\end{enumerate}

\smallskip
\noindent All the extensions above satisfy condition ${\sf Loc^f}$. 
\label{examples-local}
\end{thm}

\Proof The proof is given in Appendix~\ref{app:proof-ex-local-ext}. \qed

\subsection{Hierarchic reasoning in local theory extensions}
Let ${\mathcal T}_0 \subseteq {\mathcal T}_1 {=} {\mathcal T}_0  \cup {\mathcal K}$ 
be a local theory extension. 
To check the satisfiability 
of a set $G$ of ground clauses with respect to ${\mathcal T}_1$ we can use the 
following hierarchical procedure to reduce the problem to a satisfiability
problem in the base theory
(for details cf.\ \cite{Sofronie-cade-05}):\medskip

\begin{enumerate}[\hbox to6 pt{\hfill}]
\item\noindent{\hskip-11 pt\bf Step 1:} {\em  Use locality.} 
By the locality condition, we know that $G$ is unsatisfiable with respect to 
${\mathcal T}_1$ if and only if  ${\mathcal K}[G] \wedge G$ has no weak partial model in which 
all the subterms of ${\mathcal K}[G] \wedge G$ are defined, and whose
restriction to $\Pi_0$ is a total model of ${\mathcal T}_0$.
 
\medskip
\item\noindent{\hskip-11 pt\bf Step 2:} {\em Flattening and purification.}
As in ${\mathcal K}[G]$ and $G$ the functions in $\Sigma_1$ 
have as arguments only ground terms, ${\mathcal K}[G] \wedge G$ can be 
purified and flattened by  introducing new
constants for the arguments of the extension functions as well as for
the (sub)terms $t = f(g_1, \dots, g_n)$ 
starting with extension functions $f \in \Sigma_1$, together with new 
corresponding definitions $c_t \approx t$. The set of clauses thus 
obtained has the form ${\mathcal K}_0 \wedge G_0 \wedge D$, where $D$ is a 
set of ground unit clauses of the form $f(c_1, \dots, c_n) \approx c$, 
where $f \in \Sigma_1$ and $c_1, \dots, c_n, c$ are constants, and 
${\mathcal K}_0, G_0$ are clauses without function symbols in $\Sigma_1$.

\medskip
\item\noindent{\hskip-11 pt\bf Step 3:} {\em  Reduction to testing 
satisfiability in ${\mathcal T}_0$.} 
We reduce the problem of testing satisfiability of $G$ with respect to ${\mathcal T}_1$ 
to a satisfiability test in ${\mathcal T}_0$ as shown in Theorem~\ref{thm:hierarchic}. 
\end{enumerate}\medskip
\begin{thm}[\cite{Sofronie-cade-05}] 
Assume that ${\mathcal T}_0 \cup {\mathcal K}$ is a local extension of 
${\mathcal T}_0$ with a set ${\mathcal K}$ of clauses. 
With the notation above, the following are equivalent:

\begin{enumerate}[\em(1)]
\item ${\mathcal T}_0 \wedge {\mathcal K} \wedge G$ has a model.

\item ${\mathcal T}_0 \wedge {\mathcal K}[G] \wedge G$ has a weak 
partial model where all terms in ${\sf st({\mathcal K}, G)}$ are defined.

\item ${\mathcal T}_0 \wedge {\mathcal K}_0 \wedge G_0 \wedge D$ has a weak 
partial model with all terms in ${\sf st({\mathcal K}, G)}$ defined.

\item ${\mathcal T}_0 \wedge {\mathcal K}_0 \wedge G_0 \wedge 
{\sf Con}[D]_0$ has a (total) $\Sigma_0$-model,
  where 
\[{\sf Con}[D]_0 = \bigwedge \{ \bigwedge_{i = 1}^n c_i \approx d_i \rightarrow c \approx   d \mid 
          f(c_1, \dots, c_n) \approx c, 
          f(d_1, \dots, d_n) \approx d \in D \}\ .     
\]
is the set of instances of the congruence axioms for the functions in 
$\Sigma_1$ corresponding to the extension terms in $D$.
\end{enumerate}
\label{thm:hierarchic}
\end{thm}

\begin{exa}
\label{example-hierarchic}
Let ${\mathcal T}_1 = {\sf SLat} \cup {\sf SGc}(f, g) \cup {\sf Mon}(f, g)$
be the extension of the theory of semilattices with two monotone functions
$f, g$ satisfying the semi-Galois condition ${\sf SGc}(f, g)$. 
Consider the following ground formulae $A$, $B$ in the signature of 
${\mathcal T}_1$: 
\[A:~~ d \leq g(a) ~\wedge~ a \leq c \quad \quad 
B:~~  b \leq d ~\wedge~ f(b) \not\leq c\ .
\]
where $c$ and $d$ are shared constants. By
Theorem~\ref{examples-local}(e), ${\mathcal T}_1$ is a local extension
of the theory of semilattices. To prove that $A \wedge B
\models_{{\mathcal T}_1} \perp$ we proceed as follows:
\begin{enumerate}[\hbox to6 pt{\hfill}]
\item\noindent{\hskip-11 pt\bf Step 1:} {\em  Use locality.} 
By the locality condition, 
$A \wedge B$ is unsatisfiable with respect to 
${\sf SLat} \wedge {\sf SGc}(f, g) \wedge {\sf Mon}(f, g)$ iff 
${\sf SLat} \wedge {\sf SGc}(f, g)[A \wedge B] \wedge 
{\sf Mon}(f, g)[A \wedge B] 
\wedge A \wedge B$ has no weak partial model in which all terms in $A$ and $B$
are defined. The extension terms occurring in $A \wedge B$ are $f(b)$ and 
$g(a)$, hence:
\begin{eqnarray*}
{\sf Mon}(f, g)[A \wedge B] & = & \{ a \leq a \rightarrow g(a) \leq g(a),~~ b \leq b \rightarrow f(b) \leq f(b) \} \\
{\sf SGc}(f, g)[A \wedge B] & = & \{ b \leq g(a) \rightarrow f(b) \leq a \}  
\end{eqnarray*}

\item\noindent{\hskip-11 pt\bf Step 2:} {\em Flattening and purification.}
We purify and flatten the formula ${\sf SGc}(f, g) \wedge {\sf Mon}(f, g)$ by 
replacing the ground terms starting with $f$ and $g$ with new constants. 
The clauses are separated into a part containing definitions 
for terms starting with extension functions, $D_A \wedge D_B$, and a 
conjunction of formulae in the base signature, $A_0 \wedge B_0 \wedge {\sf SGc}_0  \wedge {\sf Mon}_0$.  

\medskip
\item\noindent{\hskip-11 pt\bf Step 3:} {\em  Reduction to testing 
satisfiability in ${\mathcal T}_0$.}
As the extension ${\sf SLat} \subseteq {\mathcal T}_1$ is local, by 
Theorem~\ref{thm:hierarchic} we know that\bigskip
\[
A\wedge B\models_{{\mathcal T}_1}\perp\qquad\text{if and only if}\qquad
\hbox to6.3 cm{\vbox to0 pt{%
\vss$A_0\wedge B_0\wedge{\sf SGc}_0\wedge{\sf Mon}_0\wedge{\sf Con}_0$\\
is unsatisfiable with respect to {\sf SLat}\,,\vss}}
\]\smallskip

\noindent where ${\sf Con}_0 = {\sf Con}[A \wedge B]_0$ consists of the
flattened form of those instances of the congruence axioms containing
only $f$- and $g$-terms which occur in $D_A$ or $D_B$, and ${\sf
  SGc}_0 \wedge {\sf Mon}_0$ consists of those instances of axioms in
${\sf SGc}(f, g) \wedge {\sf Mon}(f, g)$ containing only $f$- and
$g$-terms which occur in $D_A$ or $D_B$.

\[\begin{array}{l|ll}
\hline 
{\sf Extension} & ~~~~~~{\sf Base} \\
D_A \wedge D_B & ~A_0 \wedge B_0  \wedge {\sf SGc}_0 \wedge {\sf
  Mon}_0 \wedge 
\rlap{${\sf Con}_0$} & ~~~~~~~~~~~~~~~ \\
\hline 
a_1 \approx g(a) ~   & ~A_0 = d \leq a_1  \wedge a \leq c & {\sf SGc}_0   = b \leq a_1 \rightarrow b_1 \leq a  \\
b_1 \approx f(b)  & ~B_0 = b \leq d \wedge  b_1 \not\leq c & {\sf Con}_A \wedge {\sf Mon}_A = a \lhd a \rightarrow a_1 \lhd a_1, \lhd \in \{ \approx, \leq \}\\
& &  {\sf Con}_B \wedge {\sf Mon}_B =  b \lhd b \rightarrow b_1 \lhd b_1,~ \lhd \in \{ \approx, \leq \} \\
\hline 
\end{array}
\]\smallskip

\noindent It is easy to see that $A_0 \wedge B_0 \wedge {\sf SGc}_0
\wedge {\sf Mon}_0 \wedge {\sf Con}_0 $ is unsatisfiable with respect
to ${\mathcal T}_0$: $A_0 \wedge B_0$ entails $b \leq a_1$, together
with ${\sf SGc}_0$ this yields $b_1 \leq a$, which together with $a
\leq c$ and $b_1 \not\leq c$ leads to a contradiction.
\end{enumerate}
\end{exa}

\section{Hierarchical Interpolant Computation}
\label{hierarchic}

\noindent Let ${\mathcal T}_0 \subseteq {\mathcal T}_1 = {\mathcal T}_0 \cup {\mathcal K}$ be a
theory extension by means of a set of clauses ${\mathcal K}$. 
Assume that $A \wedge B \models_{{\mathcal T}_1} \perp$, where
$A$ and $B$ are two sets of ground clauses.  Our goal is to find a ground
{\em interpolant}, that is a ground formula $I$ containing only
constants and extension functions which are common to $A$
and $B$ such that 
$$A \models_{{\mathcal T}_1} I \quad \text{ and } \quad 
I \wedge B \models_{{\mathcal T}_1} {\perp}.$$
\noindent Flattening and purification do not influence 
the existence of ground interpolants: 
\begin{lem}
Let $A$ and $B$ be two sets of ground clauses in the signature
  $\Pi^c$. Let $A_0 \wedge D_A$ and $B_0 \wedge D_B$
  be 
obtained from $A$ resp.\ $B$  by purification and flattening. 
If $I$ is an interpolant of $(A_0 \wedge D_A) 
\wedge (B_0 \wedge D_B)$ then the formula 
${\overline I}$, obtained from $I$ by replacing,
  recursively, all newly introduced constants with the terms in the
  original signature which they represent, is an
  interpolant for $A \wedge B$.
\label{lemma:flatten}
\end{lem} 
\Proof If $I$ is an interpolant of $(A_0 \wedge D_A) 
 \wedge (B_0 \wedge D_B)$, 
 then $A_0 \wedge D_A \models_{{\mathcal T}_1} I$ and
 $B_0 \wedge D_B \wedge I \models_{{\mathcal T}_1} \perp$.
 Let ${\overline I}$ be obtained from $I$ by replacing, recursively,
 all newly introduced constants with the terms in the original
 signature which they represent. Then:
 \begin{enumerate}[(i)]
 \item $A \wedge \neg {\overline I}$
   and $A_0 \wedge D_A \wedge \neg I$ are
   equisatisfiable with respect to ${\mathcal T}_1$, 
   so $A \models_{{\mathcal T}_1} {\overline I}$.
 \item $B \wedge {\overline I}$ and
   $B_0 \wedge D_B \wedge I$ are
   equisatisfiable with respect to ${\mathcal T}_1$, so 
   $B \wedge {\overline I} \models_{{\mathcal T}_1} \perp$. \qed
 \end{enumerate} 

\medskip
\noindent 
Therefore we can restrict without loss of generality  to finding 
interpolants for the {\em purified and flattened} conjunction of formulae 
$(A_0 \wedge D_A) \wedge (B_0 \wedge D_B)$.

\medskip
We focus on interpolation in {\em local theory extensions}. 
Let ${\mathcal T}_0 \subseteq {\mathcal T}_1 = {\mathcal T}_0 \cup {\mathcal K}$ be a 
local theory extension.  
From Theorem~\ref{thm:hierarchic} we know that in such extensions 
hierarchical reasoning is possible: 
if $A$ and $B$ are sets of ground clauses in a signature $\Pi^c$, and 
$A_0 \wedge D_A$ (resp.\ $B_0 \wedge D_B$) are 
obtained from $A$ (resp.\ $B$) by purification and flattening then:  
$$(A_0 \wedge D_A) \wedge (B_0 \wedge D_B) 
\models_{{\mathcal T}_1} \perp \quad \quad \text{ if and only if } 
\quad \quad {\mathcal K}_0 \wedge 
A_0 \wedge B_0 \wedge {\sf Con}[D_A \wedge D_B]_0 \models_{{\mathcal T}_0} \perp,$$   
\noindent 
where ${\mathcal K}_0$ is obtained from 
${\mathcal K}[D_A \wedge D_B]$ by replacing the $\Sigma_1$-terms with the 
corresponding constants contained in the definitions $D_A$ and  $D_B$ and 
 
\noindent ${\sf Con}[D_A \wedge D_B]_0 = \displaystyle{ \bigwedge  \{ \bigwedge_{i = 1}^n} c_i \approx d_i   \rightarrow c \approx d 
\mid f(c_1, \dots, c_n) \approx c, f(d_1, \dots, d_n) \approx d \in D_A \cup D_B \}.$

\noindent 
In general we cannot use Theorem~\ref{thm:hierarchic} for generating 
a ground interpolant because: 
\begin{enumerate}[(i)]
\item ${\mathcal K}[D_A \wedge D_B]$ (hence also ${\mathcal K}_0$) 
may contain free variables. 

\item If some clause in ${\mathcal K}$ contains two or 
more different extension functions, it is unlikely that 
these extension functions can be separated in the interpolants. 

\item The clauses in ${\mathcal K}[D_A \wedge D_B]$ and the instances 
of congruence axioms (and therefore the clauses in ${\mathcal K}_0 \wedge {\sf Con}[D_A \wedge D_B]_0$)
may contain 
combinations of constants and extension functions from $A$ and $B$. 
\end{enumerate}

\noindent 
To avoid (i), we will need to take into account only extensions 
with sets ${\mathcal K}$ of clauses in which all variables occur below 
some extension term.
To solve (ii),  
we define a relation $\sim$ between extension functions, where 
$f \sim g$ if $f$ and $g$ occur in the same clause in ${\mathcal K}$. 
This defines an equivalence relation $\sim$ on $\Sigma_1$. 
We henceforth consider that a function $f \in \Sigma_1$ is 
common to  $A$ and $B$ if there exist $g, h \in \Sigma_1$ 
such that $f \sim g$, $f \sim h$, $g$ occurs in $A$ and $h$ occurs in $B$.
\begin{exa}
Consider the 
reduction to the base theory in Example~\ref{example-hierarchic}.
We explain the problems mentioned above.
\begin{enumerate}[\text{Ad} (i)]
\setcounter{enumi}{1}
\item As ${\sf SGc}(f, g)$ contains occurrences 
of both $f$ and $g$,  
it is not likely to find an interpolant with no occurrence of 
$f$ and $g$, even if
$g$ only occurs in $A$ and
$f$ only occurs in $B$.
We therefore assume that $f \sim g$, i.e.\ that 
both $f$ and $g$ are shared. 
\item The clause $b \leq a_1 \rightarrow b_1 \leq a$ 
of $\,{\sf SGc}_0$ 
is mixed, i.e.\ contains combinations of constants from $A$ and $B$. 
\end{enumerate}
\end{exa}

\noindent 
The idea for solving problem (iii) is presented below.

\subsection{Main Idea}

The idea of our approach is to separate mixed instances of 
axioms in ${\mathcal K}_0$, or of congruence axioms in 
${\sf Con}[D_A \wedge D_B]_0$, into 
an $A$-part and a $B$-part. 
This is, if $A \wedge B \models_{{\mathcal T}_1} \perp$ we find a set  
$T$ of $\Sigma_0 {\cup} \Sigma_1$-terms containing only constants and 
extension functions common to $A$ and $B$, such that 
${\mathcal K}[A \wedge B]$ can be separated into a part ${\mathcal K}[A, T]$ consisting of instances with extension  
terms occurring in $A$ and $T$, and a part ${\mathcal K}[B,T]$ 
containing only instances with extension terms in $B$ and $T$, such that:
\[{\mathcal K}[A,T] \wedge A_0 \wedge 
{\sf Con}[D_A \wedge D_T] \wedge 
{\mathcal K}[B,T] \wedge B_0 \wedge {\sf Con}[D_B \wedge D_T]\]
has no weak partial model where all ground terms in 
${\mathcal K}, D_A, D_B, T$  are defined.  
\begin{exa}
\label{ex-idea-interp}
Consider the conjunction $A_0 \wedge D_A \wedge B_0 \wedge D_B \wedge 
{\sf Con}[D_A \wedge D_B]_0 \wedge {\sf Mon}_0 \wedge {\sf SGc}_0$  in 
Example~\ref{example-hierarchic}. The $A$ and $B$-part share the constants 
$c$ and $d$, and no function symbols. However, as $f$ and $g$ occur together 
in ${\sf SGc}$, $f \sim g$, 
so they are considered to be all shared. (Thus, the interpolant 
is allowed to contain both $f$ and $g$.)  
We obtain a separation for the clause 
$b \leq a_1 \rightarrow b_1 \leq a$ of ${\sf SGc}_0$ as follows:
\begin{enumerate}[(i)]
\item We note that $A_0 \wedge B_0 \models b \leq a_1$. 
\item We can find an ${\sf SLat}$-term $t$ containing only shared 
constants of 
$A_0$ and $B_0$ such that $A_0 \wedge B_0 \models b \leq t \wedge t \leq a_1$.
(Indeed, such a term is $t = d$.) 
\item We show that, instead of the axiom 
$b \leq g(a) \rightarrow f(b) \leq a$, 
whose flattened form is in ${\sf SGc}_0$, we can use, without loss 
of unsatisfiability:
\begin{enumerate}[(1)]
\item an instance of the monotonicity axiom for $f$: 
$b \leq d \rightarrow f(b) \leq f(d)$, 

\item another instance of ${\sf SGc}$, namely: 
$d \leq g(a) \rightarrow f(d) \leq a$. 
\end{enumerate}
For this, we introduce a new constant $c_{f(d)}$ for $f(d)$
(its definition, $c_{f(d)} \approx f(d)$, is stored in a set $D_T$), 
and 
the corresponding instances ${\mathcal H}_{\sf sep} = {\mathcal H}^{A}_{\sf sep} 
\wedge {\mathcal H}^{B}_{\sf sep}$ 
of the congru\-ence, monotonicity and 
${\sf SGc}(f, g)$-axioms, which are now  
separated into an $A$-part 
(${\mathcal H}^{A}_{\sf sep}: d \leq a_1 \rightarrow c_{f(d)} \leq a$) and a 
$B$-part (${\mathcal H}^{B}_{\sf sep}: b \leq d \rightarrow b_1 \leq c_{f(d)}$).
We thus obtain a separated conjunction
${\overline A}_0 \wedge {\overline B}_0$  (where 
${\overline A}_0 =  {\mathcal H}^{A}_{\sf sep} \wedge A_0$ and  
${\overline B}_0 = {\mathcal H}^{B}_{\sf sep} \wedge B_0$),  
which can be proved to be unsatisfiable in 
${\mathcal T}_0 = {\sf SLat}$. 
\item To compute an interpolant in ${\sf SLat}$ for 
${\overline A}_0 \wedge {\overline B}_0$ 
note that 
${\overline A}_0$ is logically equivalent to the conjunction of unit 
literals 
\/ $d \leq a_1 ~\wedge~ a \leq c ~\wedge~ c_{f(d)} \leq a$
and ${\overline B}_0$ is logically equivalent to 
\/ $b \leq d ~\wedge~ b_1 {\not\leq} c ~\wedge~ b_1 \leq c_{f(d)}$. 
An interpolant  is 
$I_0 = c_{f(d)} \leq c$. 
\item By replacing the new constants with the 
terms they denote we obtain the interpolant 
$I = f(d) \leq c$ for $A \wedge B$. 
\end{enumerate}
\end{exa}

\noindent 
Note that in order to be able to perform in general the succession of steps in 
Example~\ref{ex-idea-interp} it is necessary that ${\mathcal K}_0$ is ground 
and the theory extension and the base theory have certain properties: 
\begin{enumerate}[(i)]
\item it always is possible to find an axiom instance 
such that all its premises are entailed by $A_0 \wedge B_0$; 
\item we can find separating terms (in the joint signature) for the entailed literals; 
\item the axioms come in pairs with corresponding monotonicity 
axioms which are then used to separate mixed rules; 
\item we can compute 
ground interpolants in ${\mathcal T}_0$.
\end{enumerate}

\smallskip
\noindent Theory extensions satisfying these conditions appear in 
a natural way in a wide variety of applications ranging from 
knowledge representation to verification. In what follows we will 
give several examples of theories with properties (i)--(iv).

\subsection{Examples of theory extensions with hierarchic interpolation}
\label{examples-sep}

We identify a class of theory extensions for which 
interpolants can be computed hierarchically (and efficiently)
using a procedure for generating 
interpolants in the base theory ${\mathcal T}_0$. 
This allows us to exploit specific properties of 
${\mathcal T}_0$ for obtaining 
simple interpolants in ${\mathcal T}_1$. 
We make the following assumptions about 
${\mathcal T}_0$:

\begin{enumerate}[\hbox to6 pt{\hfill}]
\item\noindent{\hskip-11 pt\bf Assumption 1:}\ ${\mathcal T}_0$ is {\em convex}\/ with respect to the set 
${\sf Pred}$ of all predicates (including equality $\approx$), i.e., for all conjunctions $\Gamma$ of ground 
atoms, relations $R_1, \dots, R_m \in {\sf Pred}$ and ground tuples of 
corresponding arity 
${\overline t}_1, \dots, {\overline t}_n$, if $\Gamma
  \models_{{\mathcal T}_0} \bigvee_{i = 1}^m R_i({\overline t}_i)$ 
then there exists $j \in
  \{ 1, \dots, m \}$ such that 
$\Gamma \models_{{\mathcal T}_0} R_j({\overline t}_j)$.
 
\item\noindent{\hskip-11 pt\bf Assumption 2:}\ ${\mathcal T}_0$ is
{\em $P$-interpolating} with respect to  $P \subseteq {\sf Pred}$, 
  i.e.\ for all conjunctions $A$ and $B$ of ground literals, all binary  
  predicates $R \in P$ and all
  constants $a$ and $b$ such that $a$ occurs in $A$ 
  and $b$ occurs in $B$ 
  (or vice versa), if 
$A  \wedge B \models_{{\mathcal T}_0} a R b$ then there exists a term $t$
  containing only constants common to $A$ and $B$ with 
$A \wedge B \models_{{\mathcal T}_0} a R t \wedge t R b$. 
(If we can always find a term $t$ containing only
constants common to $A$ and $B$ with 
$A \models_{{\mathcal T}_0} a R t$ and $B \models_{{\mathcal T}_0} t R b$ 
we say that ${\mathcal T}_0$ is
{\em strongly $P$-interpolating}.) 

\item\noindent{\hskip-11 pt\bf Assumption 3:}\ ${\mathcal T}_0$ has ground interpolation.
\end{enumerate}
\noindent 
Some examples of theories satisfying these properties are given below.
\begin{thm}
The following theories have ground interpolation and are convex and  
$P$-interpolating with respect to the indicated set $P$ of predicate symbols:
\begin{enumerate}[\em(1)]
\item The theory of ${\mathcal EQ}$ of pure equality without function symbols 
(for $P = \{ \approx \}$).

\item The theory ${\sf PoSet}$ of posets (for $P=\{\approx,\leq \}$).

\item Linear rational arithmetic ${\sf LI}({\mathbb Q})$ and linear
real arithmetic ${\sf LI}({\mathbb R})$ (convex with respect to $P =
\{ \approx \}$, strongly $P$-interpolating for $P = \{ \leq \}$).

\item The theories ${\sf Bool}$ of Boolean algebras, 
${\sf SLat}$ of semilattices and ${\sf DLat}$ of distributive lattices  
(strongly $P$-interpolating for $P = \{ \approx, \leq \}$).
\end{enumerate}
\label{example-assumptions-1-3}
\end{thm}
\Proof The proof is given in Appendix~\ref{app:assumptions-examples}. \qed

We make the following assumption about the extension ${\mathcal T}_1$ of 
${\mathcal T}_0$.\smallskip
\begin{enumerate}[\hbox to6 pt{\hfill}]
\item\noindent{\hskip-11 pt\bf Assumption 4:}\ ${\mathcal T}_1 =
{\mathcal T}_0 \cup {\mathcal K}$ is a local extension of ${\mathcal
T}_0$ with the property that in all clauses in ${\mathcal K}$ each
variable occurs also below some extension function.
\end{enumerate}\smallskip

\noindent For the sake of simplicity we only consider sets $A$, $B$ of
unit clauses, i.e.\ conjunctions of ground literals. This is not a
restriction, since if we can obtain interpolants for conjunctions of
ground literals then we also can construct interpolants for
conjunctions of arbitrary clauses by using standard
methods\footnote{E.g.\ in a DPLL-style procedure partial interpolants
  are generated for the unsatisfiable branches and then recombined
  using ideas of Pudl{\'a}k.} discussed e.g.\ in
\cite{McMillanProver04} or \cite{Yorsh-Musuvathi-cade-2005}.
 
\smallskip
By Lemma~\ref{lemma:flatten}, 
we can restrict  without loss of generality   to finding an
interpolant for the purified and flattened conjunction 
of unit clauses $A_0 \wedge
B_0 \wedge D_A \wedge D_B$.
By Theorem~\ref{thm:hierarchic}, 
$$A_0 \wedge D_A \wedge B_0 \wedge D_B 
\models_{{\mathcal T}_1} \perp \quad \text{ if and only if } \quad 
{\mathcal K}_0 \wedge A_0 \wedge B_0 \wedge {\sf Con}[D_A \wedge D_B]_0 
\models_{{\mathcal T}_0} 
\perp,$$
where
${\mathcal K}_0$ is obtained from 
${\mathcal K}[D_A \wedge D_B]$ by replacing the $\Sigma_1$-terms with the 
corresponding constants contained in the definitions $D_A \wedge D_B$ and 
 
\noindent 
${\sf Con}[D_A \wedge D_B]_0 = \displaystyle{ \bigwedge  \{ \bigwedge_{i = 1}^n} c_i \approx d_i   \rightarrow c \approx d 
\mid f(c_1, \dots, c_n) \approx c, f(d_1, \dots, d_n) \approx d \in D_A \cup D_B \}.$

\smallskip
\noindent 
In general, ${\sf Con}[D_A \wedge D_B]_0 = {\sf Con}^A_0 \wedge {\sf Con}^B_0 \wedge {\sf Con}_{\sf mix}$ and 
${\mathcal K}_0 = {\mathcal K}^A_0 \wedge {\mathcal K}^B_0 \wedge {\mathcal K}_{\sf mix}$,
where ${\sf Con}^A_0, {\mathcal K}^A_0$ only  contain
extension functions and 
constants which occur in $A$, ${\sf Con}^B_0, {\mathcal K}^B_0$ only  contain
extension functions and constants which occur in $B$, and 
${\sf Con}_{\sf mix}$, 
${\mathcal K}_{\sf mix}$ contain mixed clauses with constants occurring in 
both $A$ and $B$.  
Our goal is to separate ${\sf Con}_{\sf mix}$ and 
${\mathcal K}_{\sf mix}$ into an $A$-local and a $B$-local part. 
We show that, under Assumptions 1 and 2, 
${\sf Con}_{\sf mix}$ can 
always be separated, and
${\mathcal K}_{\sf mix}$ can be separated if ${\mathcal K}$ 
contains the following type of combinations of clauses:
\begin{quote}
\begin{eqnarray}
\left\{ \begin{array}{l} x_1 \, R_1 \, s_1 \wedge \dots 
\wedge x_n \, R_n \, s_n \rightarrow 
f(x_1, \dots, x_n) \, R \, g(y_1, \dots, y_n) \\ 
x_1 \, R_1 \, y_1 \wedge \dots 
\wedge x_n \, R_n \, y_n \rightarrow 
f(x_1, \dots, x_n) \, R \, f(y_1, \dots, y_n) \end{array} \right.
\label{general-form} 
\end{eqnarray}
where $n \geq 1$, $x_1, \dots, x_n$ are variables, $R_1, \dots, R_n, R$ 
are binary relations with $R_1, \dots, R_n \in P$ and $R$ transitive, 
and each $s_i$ is either a variable 
among the arguments of $g$, or a term of the form $f_i(z_1, \dots, z_k)$, 
where $f_i \in \Sigma_1$ and all the arguments of $f_i$ are 
variables occurring  among the arguments of $g$.   
\footnote{More general types of clauses, in which instead of variables 
we can consider arbitrary base terms,  can be handled if 
${\mathcal T}_0$ has a $P$-interpolation property for terms instead of 
constants. Due to space limitations, such 
extensions are not discussed here.} 
\end{quote}

\smallskip
\noindent We therefore make the following 
additional assumption 
about the theory extension ${\mathcal T}_1$:

\smallskip
\begin{enumerate}[\hbox to6 pt{\hfill}]
\item\noindent{\hskip-11 pt\bf Assumption 5:}\ 
${\mathcal T}_1 = {\mathcal T}_0 \cup {\mathcal K}$ 
is an extension of ${\mathcal T}_0$ with a set of clauses ${\mathcal K}$ 
which only contains combinations of clauses of type~(\ref{general-form}).
\end{enumerate}

\smallskip 
\begin{exa}
The following local extensions satisfy Assumptions 4 and 5: 
\begin{enumerate}[(a)]
\item Any extension with free functions (${\mathcal K} = \emptyset$). 
\item Extensions of any theory in ${\sf Ord}$ 
(cf.\ Section~\ref{sect:local-examples}) with monotone functions. 
\item Extensions of any totally-ordered theory in ${\sf Ord}$
with functions satisfying 
$${\sf SGc}(f, g_1, \dots, g_n) \wedge {\sf Mon}(f, g_1, \dots, g_n).$$
\item Extensions of theories in ${\sf Ord}$ 
with functions satisfying $${\sf SGc}(f, g_1) \wedge {\sf Mon}(f, g_1).$$
\item Extensions of theories in ${\sf Ord}$ 
with functions satisfying ${\sf Leq}(f, g) \wedge {\sf Mon}(f)$. 
\end{enumerate}
\end{exa}

\begin{rem}
If the clauses in ${\mathcal K}$ are of type~(\ref{general-form}), 
then ${\mathcal K}_0 = {\mathcal K}_0^A \wedge  {\mathcal K}_0^B \wedge {\mathcal K}_{\sf mix}$, where 
$$\begin{array}{@{}cll} 
{\mathcal K}_0^A =\!\!\! & \{ (\bigwedge_{i = 1}^n c_i R_i d_i) \rightarrow c R d \mid &\!\!\!\!  
 (\bigwedge_{i = 1}^n  x_i \, R_i \, s_i({\overline y})) \rightarrow f(x_1, \dots, x_n) \, R \, g({\overline y}) \in {\mathcal K},   \\
& & d_i \approx s_i({\overline e}) \in D_A, d \approx g({\overline e}) \in D_A, c \approx f(c_1, \dots, c_n) \in D_A \} \cup \\
& \{ (\bigwedge_{i = 1}^n c_i R_i d_i) \rightarrow c R d \mid &\!\!\!\!  
 (\bigwedge_{i = 1}^n  x_i \, R_i \, y_i) \rightarrow f(x_1, \dots, x_n) \, R \, f(y_1, \dots, y_n) \in {\mathcal K},  \\
& & d \approx f(d_1, \dots, d_n) \in D_A, c \approx f(c_1, \dots, c_n) \in D_A \},  
\end{array}$$
\noindent similarly for ${\mathcal K}_0^B$, and  
$$\begin{array}{@{}l@{}l@{}l} 
{\mathcal K}_{\sf mix} =~ & \{ \bigwedge_{i = 1}^n c_i R_i d_i \rightarrow c R d \mid~ & 
 \bigwedge_{i = 1}^n  x_i \, R_i \, s_i({\overline y}) \rightarrow f(x_1, \dots, x_n) \, R \, g({\overline y}) \in {\mathcal K}, d_i \approx s_i({\overline e}) \in D_A,   \\
& & d \approx g({\overline e}) \in D_A \backslash D_B, c \approx f(c_1, \dots, c_n) \in D_B \backslash D_A \} \cup \\[1ex]
& \{ \bigwedge_{i = 1}^n c_i R_i d_i \rightarrow c R d \mid & 
 \bigwedge_{i = 1}^n  x_i \, R_i \, y_i \rightarrow f(x_1, \dots, x_n) \, R \, f(y_1, \dots, y_n) \in {\mathcal K},   \\
& & d \approx f(d_1, \dots, d_n) \in D_A \backslash D_B, c \approx f(c_1, \dots, c_n) \in D_B \backslash D_A \} \cup \\[1ex]
& \{ \bigwedge_{i = 1}^n c_i R_i d_i \rightarrow c R d \mid & 
 \bigwedge_{i = 1}^n  x_i \, R_i \, s_i({\overline y}) \rightarrow f(x_1, \dots, x_n) \, R \, g({\overline y}) \in {\mathcal K},  d_i \approx s_i({\overline e}) \in D_B,   \\
& & d \approx g({\overline e}) \in D_B \backslash D_A, c \approx f(c_1, \dots, c_n) \in D_A \backslash D_B \} \cup \\[1ex]
& \{ \bigwedge_{i = 1}^n c_i R_i d_i \rightarrow c R d \mid & 
 \bigwedge_{i = 1}^n  x_i \, R_i \, y_i \rightarrow f(x_1, \dots, x_n) \, R \, f(y_1, \dots, y_n) \in {\mathcal K},   \\
& & d \approx f(d_1, \dots, d_n) \in D_B \backslash D_A, c \approx f(c_1, \dots, c_n) \in D_A \backslash D_B  \}. 
\end{array}$$
\noindent 
All clauses in ${\mathcal K}_0$ are of the form 
$C = \bigwedge_{i = 1}^n c_i \, R_i \, d_i {\rightarrow} c \, R \, d$, 
where $R_i \in P$, $R$ is transitive, and $c_i, d_i, c, d$ are constants. 
Moreover, the cardinality of ${\mathcal K}_0 \cup {\sf Con}[D_A \wedge D_B]_0$ 
is quadratic in the size of 
$A \wedge B$ for a fixed ${\mathcal K}$.
\end{rem}

\begin{prop} 
\label{prop:separation}
Assume that ${\mathcal T}_0$ satisfies Assumptions 1 and 2. 
Let ${\mathcal H}$ be a set of Horn clauses 
$(\bigwedge_{i = 1}^n c_i R_i d_i) \rightarrow c R d$ 
in the signature $\Pi_0^c$ (with $R$ transitive and $R_i \in P$) 
which are instances of flattened and purified clauses of 
type~(\ref{general-form}) and of congruence axioms. 
Let $A_0$ and $B_0$ be conjunctions of ground literals
in the signature $\Pi_0^c$ such that 
$A_0 \wedge B_0 \wedge {\mathcal H}  \models_{{\mathcal T}_0} \perp$. 
Then ${\mathcal H}$ can be separated into an $A$ and a $B$ part 
by replacing the set 
${\mathcal H}_{\sf mix}$ of mixed clauses

\smallskip
\noindent 
$\begin{array}{@{}lll}
{\mathcal H}_{\sf mix} & = & \{ \bigwedge_{i = 1}^n c_i R_i d_i \rightarrow c R d \in {\mathcal H} \mid c_i, c \text{ constants in } A,  d_i, d \text{ constants in } B \} \cup \\
& & \{ \bigwedge_{i = 1}^n c_i R_i d_i \rightarrow c R d \in {\mathcal H} \mid c_i, c \text{ constants in } B,  d_i, d \text{ constants in } A \} \\[0.5ex]
\end{array}$

\smallskip
\noindent with a separated set of formulae ${\mathcal H}_{\sf sep}$. 
The following hold:

\begin{enumerate}[\em(1)]
\item There exists a set $T$ of $\Sigma_0 \cup \Sigma_c$-terms containing 
only constants common to $A_0$ and $B_0$ such that 
$A_0 \wedge B_0 \wedge ({\mathcal H} \backslash {\mathcal H}_{\sf mix}) 
\wedge {\mathcal H}_{\sf sep} \models_{{\mathcal T}_0} \perp$, where 

\smallskip
\noindent 
$\begin{array}{@{}lll}
{\mathcal H}_{\sf sep} & = & \{ (\bigwedge_{i = 1}^n c_i R_i t_i \rightarrow c R c_{f(t_1, \dots, t_n)}) \wedge (\bigwedge_{i = 1}^n t_i R_i d_i \rightarrow c_{f(t_1, \dots, t_n)} R d)  \mid  \\
& & ~~ \bigwedge_{i = 1}^n c_i R_i d_i \rightarrow c R d  \in {\mathcal H}_{\sf mix}, d_i \approx s_i(e_1, \dots, e_n), d \approx g(e_1, \dots, e_n) \in D_B,  \\
& & ~~ c \approx f(c_1, \dots, c_n) \in D_A \text{ or vice versa } \} = {\mathcal H}_{\sf sep}^A \wedge {\mathcal H}_{\sf sep}^B 
\end{array}$

\smallskip 
\noindent 
and $c_{f(t_1, \dots, t_n)}$ are new constants in $\Sigma_c$ 
(considered to be common) introduced for the
corresponding terms $f(t_1, \dots, t_n)$.

\smallskip
\item $A_0 \wedge B_0 \wedge ({\mathcal H} \backslash {\mathcal H}_{\sf mix}) 
\wedge {\mathcal H}_{\sf sep}$ is logically equivalent with respect to ${\mathcal T}_0$ with 
the following separated  conjunction of ground literals: 

\smallskip
\noindent 
$\begin{array}{@{}lll}
{\overline A}_0 \wedge {\overline B}_0 = 
A_0 \,\wedge\, B_0 & \wedge &  \bigwedge \{ c R d \mid \Gamma {\rightarrow} c R d \in {\mathcal H} \backslash {\mathcal H}_{\sf mix} \} \\
& \wedge & 
\bigwedge \{  c R c_{f({\overline t})} \wedge c_{f({\overline t})} R d \mid 
(\Gamma \rightarrow c R c_{f({\overline t})}) \wedge (\Gamma \rightarrow c_{f({\overline t })} R d) \in {\mathcal H}_{\sf sep} \}.
\end{array}$

\smallskip
\item If ${\mathcal T}_0$ is strongly $P$-interpolating then the $A$-part ($B$-part) of 
$A_0 \wedge B_0 \wedge ({\mathcal H} \backslash {\mathcal H}_{\sf mix}) 
\wedge {\mathcal H}_{\sf sep} \models_{{\mathcal T}_0} \perp$ is logically equivalent 
with ${\overline A}_0$ (resp.\ ${\overline B}_0$) above. 
\end{enumerate}
\end{prop}
\Proof We prove (1) and (2) simultaneously by induction on the number of 
clauses in ${\mathcal H}$. If ${\mathcal H} = \emptyset$ then the initial 
problem is already separated into an $A$ and a $B$ part so we are done: 
we can choose $T = \emptyset$. 
Assume that ${\mathcal H}$ contains at least one clause, and that for 
every ${\mathcal H}'$ with fewer clauses and every conjunctions of 
literals $A'_0, B'_0$ with $A'_0 \wedge B'_0 \wedge {\mathcal H}' \models_{{\mathcal T}_0} \perp$, (1) and (2) hold. 

\smallskip
Let ${\mathcal D}$ be the set of all atoms $c_i R_i d_i$
occurring in premises of clauses in ${\mathcal H}$. 
As every model of $A_0 \wedge B_0 \wedge
\bigwedge_{(c R d) \in {\mathcal D}} \neg (c R d)$ is also a model for
${\mathcal H} \wedge A_0 \wedge B_0 \models_{{\mathcal T}_0} \perp$ 
and ${\mathcal H} \wedge A_0 \wedge B_0 \models_{{\mathcal T}_0} \perp$, 
$A_0 \wedge B_0 \wedge
\bigwedge_{(c R d) \in {\mathcal D}} \neg (c R d) 
\models_{{\mathcal T}_0} \perp$.  
 Let  $(A_0 \wedge B_0)^+$ be the conjunction 
of all atoms in $A_0 \wedge B_0$, and $(A_0 \wedge B_0)^-$ be the set of all 
negative literals in $A_0 \wedge B_0$. Then 
$$(A_0 \wedge B_0)^+ \models_{{\mathcal T}_0} \bigvee_{(c R d) \in {\mathcal D}} (c R d) \vee \bigvee_{\neg L \in (A_0 \wedge B_0)^-} L.$$ 
By Assumption~1, ${\mathcal T}_0$ is convex with respect to ${\sf Pred}$. 
Moreover, $(A_0 \wedge B_0)^+$ is a
conjunction of positive literals. Therefore, either 
\begin{enumerate}[(i)]
\item $(A_0 \wedge B_0)^+ \models L$ for some 
$L \in (A_0 \wedge B_0)^-$ (then 
$A_0 \wedge B_0$ is unsatisfiable and hence 
entails any atom $c_i R_i d_i$), or  
\item there exists $(c_1 R_1 d_1) \in {\mathcal D}$ such that
$(A_0 \wedge B_0)^+ \models_{{\mathcal T}_0} c_1 R_1 d_1$. 
\end{enumerate}

\smallskip
\noindent {\bf Case 1:} $A_0 \wedge B_0$ is unsatisfiable. In this case 
(1) and (2) hold for $T = \emptyset$.  

\smallskip
\noindent {\bf Case 2:} $A_0 \wedge B_0$ is satisfiable. 
Then $A_0 \wedge B_0$ is logically equivalent in ${\mathcal T}_0$ with 
$A_0 \wedge B_0 \wedge c_i R_i d_i$. 
If it is not the case that by adding $c_i R_i d_i$ all premises of some 
rule in ${\mathcal H}$ become true we repeat the procedure for 
${\mathcal D}_1 = {\mathcal D} \backslash (c_1 R_1 d_1)$:  
Again in this case $A_0 \wedge B_0 \wedge
\bigwedge_{(c R d) \in {\mathcal D}_1} \neg (c R d) 
\models_{{\mathcal T}_0} \perp$ (if it has a model then 
$A_0 \wedge B_0 \wedge {\mathcal H}$ has one), 
and as before, using convexity we infer that 
either $A_0 \wedge B_0$ is unsatisfiable (which cannot be the case) 
or there exists $c_2 R_2 d_2 \in {\mathcal D}_1$ with 
$A_0 \wedge B_0 \models_{{\mathcal T}_0} c_2 R_2 d_2$. 
We can repeat the process until all the premises of some clause 
in ${\mathcal H}$ are proved to be entailed by $A_0 \wedge B_0$. 
Let $C = \bigwedge_{i = 1}^n  c_i R_i d_i \rightarrow c R d$ be such a clause.

\smallskip
\noindent 
{\bf Case 2a.} Assume that $C$ contains only constants occurring in 
$A$ or only constants occurring in $B$.  
Then $A_0 \wedge B_0 \wedge {\mathcal H}$ is equivalent with respect to 
${\mathcal T}_0$ with 
$A_0 \wedge B_0 \wedge ({\mathcal H} \backslash C) \wedge c \approx d$. 
By the induction hypothesis for  
$A_0' \wedge B_0' = A_0 \wedge B_0 \wedge c \approx d$ and 
${\mathcal H'} = {\mathcal H} \backslash \{ C \}$,  we know that 
there exists $T'$ such that 
$A'_0 \wedge B'_0 \wedge ({\mathcal H'} \backslash {\mathcal H'}_{\sf mix}) 
\wedge {\mathcal H'}_{\sf sep} \models \perp$, and (2) holds too.
Then, for $T = T'$, 
$A'_0 \wedge B'_0 \wedge ({\mathcal H'} \backslash {\mathcal H'}_{\sf mix}) 
\wedge {\mathcal H'}_{\sf sep}$ is logically equivalent to 
$A_0 \wedge B_0 \wedge ({\mathcal H} \backslash {\mathcal H}_{\sf mix}) \wedge 
{\mathcal H}_{\sf sep}$, so (1) holds. 
In order to prove (2), note that, by definition, 
${\mathcal H}'_{\sf mix} = {\mathcal H}_{\sf mix}$ and  
${\mathcal H}'_{\sf sep} = {\mathcal H}_{\sf sep}$. 
By the induction hypothesis, 
$A'_0 \wedge B'_0 
\wedge ({\mathcal H'} \backslash {\mathcal H'}_{\sf mix}) \cup 
{\mathcal H'}_{\sf sep}$ is logically equivalent to a corresponding 
conjunction ${\overline A'_0} \wedge {\overline B'_0}$ containing 
as conjuncts all literals in $A'_0$ and $B'_0$ and all conclusions of 
rules in ${\mathcal H'} \backslash {\mathcal H'}_{\sf mix}$ and 
${\mathcal H'}_{\sf sep}$. 
On the other hand, $A'_0 \wedge B'_0$ is logically equivalent to 
$A_0 \wedge B_0 \wedge (c R d)$, where $(c R d)$ is the 
conclusion of the rule $C \in {\mathcal H} \backslash {\mathcal H}_{\sf mix}$. 
This proves (2). 

\smallskip
\noindent 
{\bf Case 2b.} Assume now that $C$ is mixed, for instance that 
$c_1, \dots, c_n, c$ are 
constants in $A$ and $d_1, \dots, d_n, d$ are constants in $B$. 
Assume that $C$ is obtained from an instance of a clause of the form 
$\bigwedge_{i = 1}^n x_i R_i s_i({\overline y}) \rightarrow 
f(x_1, \dots, x_n) R g({\overline y}).$ (The case when $C$ corresponds to 
an instance of a monotonicity axiom is similar.)
This means that there exist $c \approx f(c_1, \dots, c_n) \in D_A$ 
and $d_i \approx s_i({\overline e}), d \approx g({\overline e}) \in D_B$. 
$C$ was chosen such that for each premise $c_i R_i d_i$ of $C$, 
$A_0 \wedge B_0 \models_{{\mathcal T}_0} c_i R_i d_i$, and --
by Assumption~2 --  ${\mathcal T}_0$ is $P$-interpolating.  
Thus, there exist terms $t_1, \dots, t_n$ containing only constants 
common to $A_0$ and $B_0$ such that for all $i \in \{ 1, \dots, n \}$
\begin{eqnarray}
A_0 \wedge B_0 \models_{{\mathcal T}_0} c_i R_i t_i \wedge t_i R_i d_i. \label{interp-equation} 
\end{eqnarray}
Let $c_{f(t_1, \dots, t_n)}$ be a new constant, denoting the term 
$f(t_1, \dots, t_n)$, and let 
$$C_A  =  \bigwedge_{i = 1}^n c_i R_i t_i  {\rightarrow} 
c R c_{f(t_1, \dots, t_n)} \quad \text{ and } \quad 
C_B  =  \bigwedge_{i = 1}^n t_i R_i d_i {\rightarrow} 
c_{f(t_1, \dots, t_n)} R d.$$ 
Thus, $C_A$ corres\-ponds to the monotonicity axiom 
$\displaystyle{\bigwedge_{i = 1}^n c_i R_i t_i 
{\rightarrow} f(c_1, \dots, c_n) R f(t_1, \dots, t_n)}$, whereas  
$C_B$ corresponds to the rule $\displaystyle{\bigwedge_{i = 1}^n t_i R_i 
s_i({\overline e}) {\rightarrow} f(t_1, \dots, t_n) R g({\overline e})}$. 
As $R$ is transitive, by~(\ref{interp-equation}) the following holds:
\begin{eqnarray*}
A_0 \wedge B_0 \wedge C_A \wedge C_B & \models\!\!\!|_{{\mathcal T}_0} & A_0 \wedge B_0 \wedge (\bigwedge_{i = 1}^n c_i R_i t_i \wedge C_A) \wedge 
(\bigwedge_{i = 1}^n  t_i R_i d_i \wedge C_B) \\
& \models\!\!\!|_{{\mathcal T}_0} &   A_0 \wedge B_0 \wedge  c R c_{f(t_1, \dots, t_n)} \wedge c_{f(t_1, \dots, t_n)} R d \\
& \models_{{\mathcal T}_0} &   A_0 \wedge B_0 \wedge  c R d   
\end{eqnarray*}

~~~~~(where $\models\!\!\!|_{{\mathcal T}_0}$ stands for logical equivalence with respect to 
${\mathcal T}_0$). 

\smallskip
\noindent Hence,
$A_0 \wedge B_0 \wedge C_A \wedge C_B \wedge ({\mathcal H} \backslash C) 
 \models_{{\mathcal T}_0}    A_0 \wedge B_0 \wedge  c R d \wedge ({\mathcal H} \backslash C).$ On the other hand, 
as $A_0 \wedge B_0 \models_{{\mathcal T}_0} \bigwedge_{i = 1}^n c_i R_i d_i$,
$A_0 \wedge B_0 \wedge {\mathcal H}$ is logically equivalent with 
$ A_0 \wedge B_0 \wedge c R d \wedge ({\mathcal H} \backslash C)$, so 
$A_0 \wedge B_0 \wedge C_A \wedge C_B \wedge ({\mathcal H} \backslash C) \models_{{\mathcal T}_0} \perp$. 
By the induction hypothesis for 
$A_0 \wedge B_0 \wedge c R c_{f(t_1, \dots, t_n)} \wedge c_{f(t_1, \dots, t_n)} R d$ and ${\mathcal H}' = {\mathcal H} \backslash C$ we know that there exists a set 
$T'$ of terms such that 
$A_0 \wedge B_0 \wedge c R c_{f(t_1, \dots, t_n)} \wedge c_{f(t_1, \dots, t_n)} R d \wedge ({\mathcal H}' \backslash {\mathcal H'}_{\sf mix}) \wedge {\mathcal H}'_{\sf sep} 
\models \perp$, and also (2) holds. 
Then (1) holds for $T = T' {\cup} \{ t_1, \dots, t_n \}$. 
(2) can be proved similarly using the induction hypothesis.

\smallskip
\noindent 
(3) follows from the same induction schema taking into account 
the fact that, by strong interpolation, 
always if $A_0 \wedge B_0 \models c_i R_i d_i$ there exists $t_i$ 
(containing only 
constants common to $A_0$ and $B_0$) with 
$A_0 \models c_i R_i t_i$ and $B_0 \models t_i R_i d_i$.\footnote{In this 
case we may need to separate even pure $A$ and $B$ clauses in ${\mathcal H}$ as
in {\bf Case 2b}, in order to guarantee the separate entailment from 
$A_0$ and $B_0$.}
Then $A_0$ is logically equivalent (in ${\mathcal T}_0$) 
to $A_0 \wedge \bigwedge_{i = 1}^n c_i R_i t_i$, 
hence $A_0 \wedge C_A$ is logically equivalent to $A_0 \wedge c R c_{f(t_1, \dots, t_n)}$.  
(Similarly, $B_0$ is logically equivalent to $B_0 \wedge \bigwedge_{i = 1}^n t_i R_i d_i$, so  $B_0 \wedge C_B$ is logically equivalent to $B_0 \wedge c_{f(t_1, \dots, t_n)} R d$.) By using the induction hypothesis, (3) follows easily.
\qed

\medskip
\noindent 
An immediate consequence of Proposition~\ref{prop:separation} is 
Proposition~\ref{thm:separate}.
\begin{prop}
\label{thm:separate}
Assume ${\mathcal T}_0$ satisfies Assumptions 1 and 2, the extension
${\mathcal T}_0 \subseteq {\mathcal T}_0 \cup {\mathcal K}$ satisfies 
Assumptions 4 and 5, and    
${\mathcal K}_0 \wedge A_0 \wedge B_0 \wedge {\sf Con}[D_A \wedge D_B]_0 
\models_{{\mathcal T}_0} {\perp}$.
Then there exists a set $T$ of $\Sigma_0 \cup \Sigma_c$-terms containing 
only constants common to $A_0$ and $B_0$ such that 
(if  ${\sf Con}^{D}_0 = {\sf Con}^{DA}_0 \wedge {\sf Con}^{DB}_0 {=} 
{{\sf Con}_0}_{\sf sep}$ and 
${\mathcal K}^D_0 = {\mathcal K}^{DA}_0 \wedge {\mathcal K}^{DB}_0 {=} 
{{\mathcal K}_0}_{\sf sep}$):
\begin{eqnarray}
{\mathcal K}^A_0 \wedge {\mathcal K}^B_0 \wedge {\mathcal K}^D_0 \wedge A_0 
\wedge B_0 \wedge {\sf Con}^A_0 \wedge {\sf Con}^B_0 \wedge 
{\sf Con}^D_0 \models_{{\mathcal T}_0} {\perp}.\label{end}
\end{eqnarray} 
\noindent 
As before, $\Sigma_c$ contains the new constants 
$c_{f(t_1, \dots, t_n)}$, considered to be common to $A_0$ and $B_0$,  
introduced for terms $f(t_1, \dots, t_n)$, with 
$t_1, \dots, t_n \in T$.
\end{prop}
\Proof If ${\mathcal K}$ only contains 
combinations of clauses of type~(\ref{general-form}) then 
all clauses in ${\mathcal K}_0 \wedge {\sf Con}[D_A \wedge D_B]_0$ 
satisfy the restrictions  
on ${\mathcal H}$ in Proposition~\ref{prop:separation}. Thus 
Proposition~\ref{prop:separation} holds for 
${\mathcal H} = {\mathcal K}_0 \wedge {\sf Con}[D_A \wedge D_B]_0$. 
Therefore there exists a 
set $T$ of $\Sigma_0 \cup \Sigma_c$-terms containing 
only constants common to $A_0$ and $B_0$ such that 
$A_0 \wedge B_0 \wedge ({\mathcal H} \backslash {\mathcal H}_{\sf mix}) 
\wedge {\mathcal H}_{\sf sep}$. The statement of the theorem uses the 
description of ${\mathcal H} \backslash {\mathcal H}_{\sf mix}$, denoted 
before by ${\mathcal K}_0^A \wedge {\mathcal K}_0^B$, as well as of 
${\mathcal H}_{\sf sep}$  as 
${\mathcal K}_0^{DA} \wedge {\mathcal K}_0^{DB}  \wedge {\sf Con}_0^{DA} 
\wedge {\sf Con}_0^{DB}$. \qed
\begin{cor}
Assume that the extension 
${\mathcal T}_0 \subseteq {\mathcal T}_0 \cup {\mathcal K}$ 
satisfies Assumptions 1--5, and that 
${\mathcal K}_0 \wedge A_0 \wedge B_0 \wedge {\sf Con}[D_A \wedge D_B]_0 
{\models}_{{\mathcal T}_0} {\perp}$. With the notation in 
Proposition~\ref{thm:separate} the following holds:

\begin{enumerate}[\em(1)]
\item There exists a $\Pi_0^c$-formula $I_0$ containing only constants
common to $A_0$, $B_0$ with 
${\mathcal K}^{A}_0 \wedge {\mathcal K}^{DA}_0 \wedge A_0 \wedge 
{\sf Con}^{A}_0 \wedge {\sf Con}^{DA}_0 {\models}_{{\mathcal T}_0} I_0$ and  
${\mathcal K}^{B}_0 \wedge {\mathcal K}^{DB}_0 \wedge B_0 \wedge 
{\sf Con}^{B}_0 \wedge {\sf Con}^{DB}_0 \wedge I_0 
{\models}_{{\mathcal T}_0} {\perp}.$

\item
There exists a ground $\Pi^c$-formula $I$ containing only constants
and function symbols which occur both in $A$ and $B$ such that
$A \models_{{\mathcal T}_1} I$ and 
$B \wedge I \models_{{\mathcal T}_1} \perp.$
\end{enumerate}
\label{thm:interp}
\end{cor}
\Proof\hfill
\begin{enumerate}[(1)] 
\item is a direct consequence of Proposition~\ref{thm:separate}, 
since ${\mathcal K}^{A}_0, {\mathcal K}^{AD}_0, 
{\mathcal K}^{B}_0, {\mathcal K}^{BD}_0$ are ground and we assumed that 
${\mathcal T}_0$ has ground interpolation. 

\smallskip
\item Let $I$ be obtained from $I_0$
by recursively replacing each constant $c_t$ 
introduced in the separation process with the term $t$. 
We show that $I$ is an interpolant of 
$(A_0 \wedge D_A) \wedge (B_0 \wedge D_B)$ with respect to ${\mathcal T}_1$, 
i.e.\ that (i) $A_0 \wedge D_A \models_{{\mathcal T}_1} I$ and 
(ii) $I \wedge B_0 \wedge D_B \models_{{\mathcal T}_1} \perp$. 

\smallskip
\begin{enumerate}[(i)]
\item Let $(M, v)$ be a ${\mathcal T}_1$-model that satisfies $A_0 \wedge D_A$. 
Being a model of ${\mathcal T}_1$, $(M, v)$ satisfies all instances of 
the axioms in ${\mathcal K}$ and of the congruence axioms in 
${\mathcal K}^{A}_0 \wedge {\mathcal K}^{DA}_0 
\wedge {\sf Con}^{A}_0 \wedge {\sf Con}^{DA}_0$ 
(and similarly for the $B$ part). 
Thus, the restriction $(M_{|\Pi_0}, v)$ of $(M, v)$ to the base theory 
satisfies ${\mathcal K}^{A}_0 \wedge {\mathcal K}^{DA}_0 \wedge A_0 \wedge 
{\sf Con}^{A}_0 \wedge {\sf Con}^{DA}_0$, hence also $I_0$. 
We thus proved that $A_0 \wedge D_A \models_{{\mathcal T}_1} I_0 \wedge D_A$. 
It is easy to see that $I_0 \wedge D_A \models_{{\mathcal T}_1} I$. 

\smallskip
\item Assume that 
$I \wedge  B_0 \wedge D_B$ has a ${\mathcal T}_1$-model 
$(M, v)$. Then $(M, v) \models I_0 \wedge B_0 \wedge D_B$, so 
its reduct to $\Pi_0$ is a model of ${\mathcal T}_0$ and of 
${\mathcal K}^{B}_0 \wedge {\mathcal K}^{DB}_0 \wedge B_0 \wedge 
{\sf Con}^{B}_0 \wedge {\sf Con}^{DB}_0 \wedge I_0$. 
This contradicts the fact that 
the set of clauses above is unsatisfiable with respect to ${\mathcal T}_0$. Thus, $
I \wedge B_0 \wedge D_B \models_{{\mathcal T}_1} \perp$.\qed 
\end{enumerate}
\end{enumerate}

\section{A procedure for hierarchical interpolation}
\label{procedure}
\noindent 
We obtain a procedure for computing interpolants for $A \wedge B$ 
described in Figure~\ref{fig:procedure}. 

\begin{figure}[h]
{\small 
\noindent \begin{tabular}{l}
$\mbox{\hspace{15cm}}$  \\
\hline 
\end{tabular}

\smallskip 
\noindent 
\begin{tabular}{@{}ll}
{\em Given:} & Local extension ${\mathcal T}_0 \subseteq {\mathcal T}_1 = {\mathcal T}_0 \cup {\mathcal K}$ which satisfies Assumptions~1--5; \\
             & Conjunctions $A$ and $B$ of literals over the signature of ${\mathcal T}_1$   such that $A \wedge B \models_{{\mathcal T}_1} \perp$ \\[2ex]
{\em Task:} & Find an interpolant for $A \wedge B$, i.e.\ a formula $I$ with $A \models_{{\mathcal T}_1} I$ and $I \wedge B \models_{{\mathcal T}_1} \perp$. \\[2ex]

{\em Method:} & \\
\end{tabular}

\smallskip
\begin{description}
\item[Step 1] {\em Purify.} \\

\noindent Using locality, flattening and purification we obtain a 
set ${\mathcal H} \wedge A_0 \wedge B_0$ of formulae in the base theory, where 
${\mathcal H} = {\mathcal K}_0 \wedge {\sf Con}[D_A \wedge D_B]_0$.

\noindent Let $\Delta := {\sf T}$.

\smallskip
\item[Step 2] {\em Reduction to an interpolation problem in the base theory}. \hspace{2mm}\\

\noindent {\em Repeat as long as possible:} \\
\noindent ~~Let $C {\in} {\mathcal H}$ whose premise is entailed by 
$A_0 \wedge B_0 \wedge \Delta$. \\
\noindent ~~\begin{tabular}{@{}l@{}l}
If $C$ is mixed,~ & compute terms $t_i$ which separate the premises in $C$, 
and separate \\
& the clause into an instance $C_1$ of monotonicity and 
an instance $C_2$ \\
& of a clause in ${\mathcal K}$ as in the proof of Case 2b in Prop.~\ref{prop:separation}. \\  
& Remove $C$ from ${\mathcal H}$, and add $C_1, C_2$ to ${\mathcal H}_{\sf sep}$ 
and their conclusions  to $\Delta$. 
\end{tabular}

\noindent ~~Otherwise move $C$ to ${\mathcal H}_{\sf sep}$ and add its conclusion
to $\Delta$.

\smallskip
\item[Step 3] {\em Interpolation in the base theory.} \\
\noindent 
Compute an interpolant $I_0$ in ${\mathcal T}_0$ for the separated formula 
${\overline A}_0 \wedge {\overline B}_0$ (logically equivalent to 
$A_0 \wedge B_0 \wedge ({\mathcal H} \backslash {\mathcal H}_{\sf mix}) \wedge 
{\mathcal H}_{\sf sep}$) obtained this way. 

\smallskip
\item[Step 4] {\em Construct interpolant for the initial problem.}\\
\noindent 
Construct an interpolant $I$ in ${\mathcal T}_1$ 
from $I_0$
by recursively replacing each constant $c_t$ 
introduced in the separation process with the term $t$, 
as explained in Corollary~\ref{thm:interp}(2). 
\end{description}

\noindent \begin{tabular}{l}
$\mbox{\hspace{15cm}}$  \\
\hline 
\end{tabular}
}
\caption{Procedure for hierarchical interpolant computation}
\label{fig:procedure}
\end{figure}

\begin{lem}\label{thm:very-local} 
Assume that the cycle in Step 2 of the procedure described in 
Figure~\ref{fig:procedure} stops after processing all mixed clauses 
in ${\mathcal H}_{\sf mix}$ and moving their separated form  into 
the set ${\mathcal H}_{\sf sep}$. The following are equivalent:
\begin{enumerate}[\em(1)]
\item $A_0 \wedge D_A \wedge B_0
  \wedge D_B \models_{{{\mathcal T}_1}} \perp$.

\item $A_0 \wedge B_0 \wedge ({\mathcal H} \backslash {\mathcal H}_{\sf mix}) \wedge {\mathcal H}_{\sf sep} \models_{{\mathcal T}_0} \perp$. 
\end{enumerate}
\end{lem}
\Proof $(1) {\Rightarrow} (2)$ is a consequence of 
Theorems~\ref{thm:hierarchic}
and Proposition~\ref{thm:separate}.
As the conjunction in (2) corresponds to a subset of instances of 
${\mathcal K} \wedge A_0 \wedge D_A \wedge B_0 \wedge D_B$, 
$(2)$ implies $(1)$. \qed

\medskip
\noindent 
{\bf Note:}  
If ${\mathcal K}_0 \wedge A_0 \wedge B_0 \wedge {\sf Con}[D_A \wedge D_B]_0 
\models_{{{\mathcal T}_0}}\perp$  then no matter which  
terms are chosen for separating mixed clauses in 
${\sf Con}[D_A \wedge D_B]_0 \wedge {\mathcal K}_0$, 
we obtain a separated conjunction of clauses unsatisfiable with respect to 
${\mathcal T}_0$.
Lemma~\ref{thm:very-local} shows that if the set of clauses obtained when 
the procedure stops is satisfiable then $A \wedge B$ was 
satisfiable, and conversely, so the procedure 
can be used to test satisfiability and to compute 
interpolants at the same time. (However, it is more efficient to first test  
$A \wedge B \models_{{\mathcal T}_1} \perp$.) 

\begin{thm}
Let ${\mathcal T}_0$ be a theory with the following properties: 
\begin{enumerate}[\hbox to6 pt{\hfill}]
\item\noindent{\hskip-0 pt\bf Assumption 1:}\ ${\mathcal T}_0$ is
{\em convex}\/ with respect to the set ${\sf Pred}$ (including
equality $\approx$);
\item\noindent{\hskip-0 pt\bf Assumption 2:}\ ${\mathcal T}_0$ is {\em $P$-interpolating} with respect to a subset $P \subseteq {\sf Pred}$
and the separating terms $t_i$ can be effectively computed;
and 
\item\noindent{\hskip-0 pt\bf Assumption 3:}\ ${\mathcal T}_0$ has ground interpolation 
\end{enumerate}
(note that we assume, in particular, that 
${\mathcal T}_0$ satisfies a stronger form of Assumption 2). 
Assume that the extension 
${\mathcal T}_1 = {\mathcal T}_0 \cup {\mathcal K}$ of 
${\mathcal T}_0$ has the following properties: 
\begin{enumerate}[\hbox to6 pt{\hfill}]
\item\noindent{\hskip-0 pt\bf Assumption 4:}\ ${\mathcal T}_1$ is a local extension of ${\mathcal T}_0$; and 
\item\noindent{\hskip-0 pt\bf Assumption 5:}\ ${\mathcal K}$ consists
of the following type of combinations of clauses:
\begin{eqnarray*}
\left\{ \begin{array}{l} x_1 \, R_1 \, s_1 \wedge \dots 
\wedge x_n \, R_n \, s_n \rightarrow 
f(x_1, \dots, x_n) \, R \, g(y_1, \dots, y_n) \\ 
x_1 \, R_1 \, y_1 \wedge \dots 
\wedge x_n \, R_n \, y_n \rightarrow 
f(x_1, \dots, x_n) \, R \, f(y_1, \dots, y_n) \end{array} \right.
\end{eqnarray*}
where $n \geq 1$, $x_1, \dots, x_n$ are variables, $R_1, \dots, R_n, R$ 
are binary relations, $R_1, \dots, R_n \in P$, $R$ is transitive, 
and each $s_i$ is either a variable 
among the arguments of $g$, or a term of the form $f_i(z_1, \dots, z_k)$, 
where $f_i \in \Sigma_1$ and all the arguments of $f_i$ are 
variables occurring  among the arguments of $g$  (i.e.\ combinations of 
clauses of type~(\ref{general-form})).  
\end{enumerate}
For every conjunction $A \wedge B$ of ground unit clauses in the signature 
$\Pi^c$ of ${\mathcal T}_1$ (possibly containing additional constants)
with $A \wedge B \models_{{\mathcal T}_1} \perp$ the procedure for 
hierarchical interpolation terminates and it computes 
an interpolant $I$ for $A \wedge B$. 
\end{thm}
\Proof To prove termination note that at every execution of the loop in Step 2, the number of 
mixed clauses decreases. All entailment tests in Step 2 are decidable 
(their complexity is discussed separately). 
By Assumption (2'), terms $t_i$ which separate the premises can be 
computed in finite time. This shows that Step 2 terminates. 
Termination of Steps 1, 3 and 4 is immediate.

We now prove correctness. We know that 
$A \wedge B \models_{{\mathcal T}_1} \perp$, so 
$A_0 \wedge D_A \wedge B_0 \wedge D_B \models_{{\mathcal T}_1} \perp$. 
Hence, by Lemma~\ref{thm:very-local}, 
when the cycle in Step 2 of the procedure terminates 
replacing the set of clauses ${\mathcal H}_{\sf mix}$ 
with ${\mathcal H}_{\sf sep}$, then  
$A_0 \wedge B_0 \wedge ({\mathcal H} \backslash {\mathcal H}_{\sf mix}) \wedge {\mathcal H}_{\sf sep} \models_{{\mathcal T}_0} \perp$. 
By construction, at termination 
${\mathcal H} \backslash {\mathcal H}_{\sf mix} \wedge {\mathcal H}_{\sf sep}$ 
contains only  pure (unmixed) clauses. We can use the alternative form of 
${\mathcal H} \backslash {\mathcal H}_{\sf mix}$, denoted 
before by ${\mathcal K}_0^A \wedge {\mathcal K}_0^B$, as well as of 
${\mathcal H}_{\sf sep}$  as ${\mathcal K}_0^{DA} \wedge {\mathcal K}_0^{DB} 
\wedge {\sf Con}_0^{DA} \wedge {\sf Con}_0^{DB}$. 
In Step 3 an interpolant 
$I_0$ containing only constants
common to $A_0$, $B_0$ with 
${\mathcal K}^{A}_0 \wedge {\mathcal K}^{DA}_0 \wedge A_0 \wedge 
{\sf Con}^{A}_0 \wedge {\sf Con}^{DA}_0 {\models}_{{\mathcal T}_0} I_0$ 
and  
${\mathcal K}^{B}_0 \wedge {\mathcal K}^{DB}_0 \wedge B_0 \wedge 
{\sf Con}^{B}_0 \wedge {\sf Con}^{DB}_0 \wedge I_0 
{\models}_{{\mathcal T}_0} {\perp}$ is computed.
In Step 4, a ground $\Pi^c$-formula $I$ 
containing only constants
and function symbols which occur both in $A$ and $B$ such that
$A \models_{{\mathcal T}_1} I$ and 
$B \wedge I \models_{{\mathcal T}_1} \perp$ is constructed starting from $I$ 
as explained in Corollary~\ref{thm:interp}. 
This is the interpolant of $A \wedge B$.
\qed

\medskip
\noindent 
{\bf Complexity:} Assume that in ${\mathcal T}_0$ for a formula of 
length $n$: 
\begin{enumerate}[(a)]
\item interpolants can be computed in time $g(n)$, 
\item $P$-interpolating terms can be computed in time $h(n)$, 
\item entailment can be checked in time $k(n)$. 
\end{enumerate}
The size $n$ of the set of clauses obtained after the preprocessing
phase is quadratic in the size of the input. Under the assumptions
(a), (b), (c) above the procedure above computes an interpolant in
time of order $n \cdot (k(n) {+} h(n)) {+} g(n)$.
\begin{rem}
If ${\mathcal T}_0$ satisfies Assumptions 1 and 3 at the beginning of 
Section~\ref{examples-sep}. 
and is strongly $P$-interpo\-la\-ting, the procedure above can be 
changed (according to the proof of Proposition~\ref{prop:separation}(3)) 
to separate {\em all}\/ clauses in ${\mathcal H}$ 
and store the conclusions of the separated clauses 
in $\Delta = \Delta_A \cup \Delta_B$.
If ${\mathcal K}_0 \wedge A_0 \wedge B_0 \wedge {\sf Con}[D_A \wedge D_B]_0 
{\models}_{{\mathcal T}_0} {\perp}$ then there exists a set $T$ 
of $\Sigma_0 \cup \Sigma_c$-terms containing 
only constants common to $A_0$ and $B_0$, and 
  common new constants in a set $\Sigma_c$ such that the
  terms in $T$ can be used to separate ${\sf Con}[D_A \wedge D_B]_0 
  \cup {\mathcal K}_0 $ into 
   ${\mathcal H}_{\sf sep} = ({\mathcal K}^{DA}_0 \wedge {\sf Con}^{DA}_0) \wedge 
({\mathcal K}^{DB}_0 \wedge {\sf Con}^{DB}_0)$, where:
  $$\begin{array}{lll}
    {\mathcal H}_{\sf sep} & = & \{ (\bigwedge_{i = 1}^n c_i R_i t_i \rightarrow c R c_{f(t_1, \dots, t_n)}) \wedge (\bigwedge_{i = 1}^n t_i R_i d_i \rightarrow c_{f(t_1, \dots, t_n)} R d)  \mid  \\
    & & ~~ \bigwedge_{i = 1}^n c_i R_i d_i \rightarrow c R d \in {\sf Con}_0 \cup {\mathcal K}_0 \} = ({\mathcal K}^{DA}_0 \wedge {\sf Con}^{DA}_0) \wedge ({\mathcal K}^{DB}_0 \wedge {\sf Con}^{DB}_0)
  \end{array}$$
\noindent 
such that for each premise $c_i R_i d_i$ of a rule in ${\sf Con}[D_A \wedge D_B]_0 \cup {\mathcal K}_0$,
at some step in the procedure 
$A_0 \wedge B_0 \wedge \Delta_A \wedge \Delta_B \models c_i R_i d_i$ 
and there exists $t_i \in T$ such
  that $A_0 \wedge \Delta_A \models c_i R_i t_i$ and 
  $B_0 \wedge \Delta_B \models t_i R_i d_i$. In this
  case $A_0 \wedge {\mathcal K}^{DA}_0 \wedge {\sf Con}^{DA}_0$ is logically equivalent to
  ${\overline A}_0$, and $B_0 \wedge {\mathcal K}^{DB}_0 \wedge {\sf Con}^{DB}_0$ 
  is logically
  equivalent to ${\overline B}_0$, where ${\overline A}_0$,
  ${\overline B}_0$ are the following conjunctions of literals:
$$\begin{array}{lll} {\overline A}_0 & = & A_0 \wedge
    \bigwedge \{ c R c_{f({\overline t})} \mid \text{ conclusion of
      some clause }
    (\Gamma \rightarrow c R c_{f({\overline t})}) \in  {\mathcal K}^{DA}_0 
    \cup {\sf Con}^{DA}_0\} \\
    {\overline B}_0 & = & B_0 \wedge \bigwedge \{ c_{f({\overline t})}
    R d \mid \text{ conclusion of some clause } (\Gamma \rightarrow
    c_{f({\overline t })} R d) \in {\mathcal K}^{DB}_0 \cup 
    {\sf Con}^{DB}_0 \}.
  \end{array}$$
\noindent 
Thus, if for instance in ${\mathcal T}_0$ 
interpolants for conjunctions 
of ground literals are always 
again conjunctions of ground literals,  the same is 
also true in the extension. 
\label{remark-interpolants}
\end{rem}
\begin{exa} 
The following theory extensions have ground interpolation: 
\begin{enumerate}[(a)]
\item Extensions of any theory in 
Theorem~\ref{example-assumptions-1-3}(1)--(4) with free function symbols.

\item Extensions of the theories in 
Theorem~\ref{example-assumptions-1-3}(2),(4) with monotone functions.

\item Extensions of the theories in 
Theorem~\ref{example-assumptions-1-3}(2),(4) with 
${\sf Leq}(f,g) \wedge {\sf Mon}(f)$.

\item Extensions of the theories in 
Theorem~\ref{example-assumptions-1-3}(2),(4) with 
${\sf SGc}(f, g_1) \wedge {\sf Mon}(f, g_1)$.

\item  Extensions of any theory in 
Theorem~\ref{example-assumptions-1-3}(1)--(4) with 
${\sf Bound}^t(f)$ or ${\sf GBound}^t(f)$ 
(where $t$ is a term and $\phi$ a set of literals in the 
base theory). 

\item Extensions of the theories in 
Theorem~\ref{example-assumptions-1-3}(2),(4) with 
${\sf Mon}(f) \wedge {\sf Bound}^t(f)$, if $t$ is monotone in its variables.

\item  ${\mathbb R} \cup ({\sf L}_f^{\lambda})$, 
the extension of the theory of reals  with a unary function which is 
$\lambda$-Lipschitz in a point $x_0$, 
where $({\sf L}_f^{\lambda})$ is 
$\forall x ~ | f(x) - f(x_0) | \leq \lambda \cdot | x - x_0|$.
\end{enumerate}
\end{exa}
\Proof (a)--(d) are direct consequences of Corollary~\ref{thm:interp}, 
since all sets of 
extension clauses are of type~(\ref{general-form}). 
For extensions of linear arithmetic 
note that due to  the totality of 
$\leq$ we can always assume that $A$ and $B$ are positive, so 
 convexity with respect to $\approx$ is sufficient 
(cf. proof of Proposition~\ref{prop:separation}). Also, in \cite{RybalchenkoSofronie06} we show that being $P$-interpolating with respect to $\leq$ is sufficient in this case.
(e)--(g) follow from Corollary~\ref{thm:interp} and the fact that if each 
clause in
${\mathcal K}$ contains only one occurrence of an extension function, no 
mixed instances can be generated when computing ${\mathcal K}[A \wedge B]$. \qed
\section{Applications}
\label{appl}
\subsection{Modular reasoning in local combinations of theories}
\label{reasoning-local-comb}
Let 
${\mathcal T}_i = {\mathcal T}_0 \cup {\mathcal K}_i, i = 1, 2$ 
be local extensions of a theory ${\mathcal T}_0$ with signature  
$\Pi_0 = (\Sigma_0, {\sf Pred})$, where $\Sigma_0 = \Sigma_1 \cap \Sigma_2$. 
Assume that 
(a) all variables in ${\mathcal K}_i$ occur below some extension function,
(b) the extension ${\mathcal T}_0 \subseteq 
{\mathcal T}_0 \cup {\mathcal K}_1 \cup {\mathcal K}_2$ 
is local\footnote{If ${\mathcal T}_0$ is a $\forall\exists$ theory then 
(b) is implied by (a) and the locality of ${\mathcal T}_1, {\mathcal T}_2$ 
\cite{Sofronie-frocos07}.}, and  
(c) ${\mathcal T}_0$ has ground interpolation. 

\smallskip

Let $G$ be a set of ground clauses in the signature $\Pi^c = (\Sigma_0
\cup \Sigma_1 \cup \Sigma_2 \cup \Sigma_c, {\sf Pred})$.  $G$ can be
flattened and purified, so we assume without loss of generality that
$G = G_1 \wedge G_2$, where $G_1, G_2$ are flat and linear sets of
clauses in the signatures $\Pi_1, \Pi_2$ respectively, i.e. for $i =
1, 2$, $G_i = G^0_i \wedge G_0 \wedge D_i$, where $G^0_i$ and $G_0$
are clauses in the base theory and $D_i$ conjunctions of unit clauses
of the form $f(c_1, \dots, c_n) = c, f \in \Sigma_i \backslash
\Sigma_0$.
\begin{thm}
With the notations above, 
assume that  
$G_1 \wedge G_2 \models_{{\mathcal T}_1 \cup {\mathcal T}_2} \perp$.
Then there exists a ground formula $I$, containing only constants shared 
by $G_1$ and $G_2$, with $G_1 \models_{{\mathcal T}_1 {\cup} {\mathcal T}_2} I$ and $I \wedge G_2 \models_{{\mathcal T}_1 {\cup} {\mathcal T}_2} \perp$.  
\end{thm}
\Proof By Theorem~\ref{thm:hierarchic}, the following are equivalent:

\begin{enumerate}[(1)]
\item ${\mathcal T}_0 \cup {\mathcal K}_1 \cup {\mathcal K}_2 \cup (G^0_1 \wedge G_0 \wedge D_1) \wedge (G^0_2 \wedge G_0 \wedge D_2) \models \perp$, 
\item ${\mathcal T}_0 \cup {\mathcal K}_1[G_1] \wedge {\mathcal K}_2[G_2] \wedge 
 (G^0_1 \wedge G_0 \wedge D_1) \wedge (G^0_2 \wedge G_0 \wedge D_2) 
\models \perp$, 
\item ${\mathcal K}^0_1 \wedge {\mathcal K}^0_2 \wedge (G^0_1 \wedge G_0) \wedge (G^0_2 \wedge G_0) \wedge {\sf Con}_1 \wedge {\sf Con}_2 
\models_{{\mathcal T}_0} \perp$,  
where, for $j = 1, 2,$ 
\end{enumerate}

${\sf Con}_j = \displaystyle{\bigwedge \{ \bigwedge_{i = 1}^n c_i \approx d_i \rightarrow c \approx d \mid 
f(c_1, \dots, c_n) \approx c, f(d_1, \dots, d_n)\approx d \in D_j \}},$\\ 
and ${\mathcal K}^0_i$ is the formula 
obtained from ${\mathcal K}_i[G_i]$ after purification and flattening, taking into 
account the definitions from $D_i$. 
Let $A = {\mathcal K}^0_1 \wedge (G^0_1 \wedge G_0) \wedge {\sf Con}_1$ and 
$B = {\mathcal K}^0_2 \wedge (G^0_2 \wedge G_0) \wedge {\sf Con}_2$. 
By assumption (a), 
$A$ and $B$ are both ground. As $A$ and 
$B$ have no extension function symbols in common and only  share
the constants which $G_1$ and $G_2$ share, 
there exists an interpolant $I_0$ in the signature $\Pi_0$, 
containing only $\Sigma_0$-function symbols and only constants shared by 
$G_1, G_2$, such that $A \models_{{\mathcal T}_0} I_0$ and 
$B \wedge I_0 \models_{{\mathcal T}_0} \perp$. An interpolant for 
$G_1 \wedge G_2$ with respect to ${\mathcal T}_1$ can now be obtained by replacing 
the newly introduced constants by the terms they replaced. \qed 

\smallskip
\noindent 
By Remark~\ref{remark-interpolants}, if ${\mathcal T}_0$ is strongly 
$P$-interpolating and has equational 
interpolation then $I$ is a conjunction of literals, so 
for modularly proving $G_1 \wedge G_2 \models_{{\mathcal T}_1} \perp$ 
only conjunctions of ground literals containing constants 
shared by $G_1, G_2$ need to be exchanged between specialized provers for 
${\mathcal T}_1$ and ${\mathcal T}_2$.

\subsection{Terminological Databases}
\label{appl-knowledge}
Consider the combination of databases in 
Section~\ref{motivation-knowledge}. 
We prove that 
\begin{eqnarray}
\Gamma_0 \wedge ({\sf T}_1 \wedge \Gamma_1) \wedge ({\sf T}_2 \wedge \Gamma_2) \models_{\mathcal T} \perp. \label{incons-sl}
\end{eqnarray}
\noindent 
where ${\mathcal T}$ is the extension 
${\sf SLat} {\cup} {\bigcup}_{f \in {\sf R}_1 {\cup} {\sf R}_2} {\sf Mon}(f)$ of the theory of 
semilattices with 0 and 
monotone functions corresponding to the r{\^o}les in ${\sf R}_1 {\cup} {\sf R_2}$, where: 

\vspace{-3mm}
\begin{eqnarray*}
{\Gamma}_0 &\!\!\!=\!\!\!& \{ {\sf organic} \wedge {\sf inorganic} \approx 0,   \quad 
                      {\sf organic} \leq {\sf substance},  \quad  {\sf inorganic} \leq {\sf substance} \}  \\[1ex]
{\sf T}_1 &\!\!\!=\!\!\!& \{ {\sf cat}\text{-}{\sf oxydation} \approx {\sf substance} \wedge {\sf catalyzes}({\sf oxydation}) \} \quad \quad \quad \quad \quad \quad \quad \quad \quad \quad \quad \quad \quad \quad \quad \\
\Gamma_1 &\!\!\!=\!\!\!& \{ 
                {\sf reaction} \leq {\sf oxydation}, 
                {\sf cat}\text{-}{\sf oxydation} \leq {\sf inorganic},  
                {\sf cat}\text{-}{\sf oxydation} {\not\approx} 0 \}, \Gamma_2 = \{ {\sf enzyme} \not\approx 0 \}\\[1ex]
{\sf T}_2 &\!\!\!=\!\!\!& \{ {\sf reaction} \approx {\sf process} \wedge {\sf produces}({\sf substance}), \quad  {\sf enzyme} \approx {\sf organic} \wedge {\sf catalyzes}({\sf reaction}) \} 
\end{eqnarray*} 

\smallskip
\noindent 
In order to find the mistake 
we look for an  
explanation for the inconsistency in the joint language of the two databases. 
Based on results on hierarchical reasoning in extensions of 
theories in \cite{Sofronie-cade-05} we can show that if we purify the 
problem by introducing definitions for the terms starting with an extension 
role symbol we can reduce the satisfiability test to a satisfiability test 
in the base theory. 
Thus,~(\ref{incons-sl}) is 
equivalent to the unsatisfiability of a set of clauses over the theory of 
semilattices, namely: $C_{T_0} \wedge C_{T_1} \wedge 
C_{\Gamma_1} \wedge C_{T_2} \wedge C_{\Gamma_2} \wedge N$ where $C_{T_i}$ and
$C_{\Gamma_i}$ are as
in the table below (the shared symbols are underlined): 

{\small 
\medskip
\noindent 
$\begin{array}{|l||l||l|l|}
\hline 
 & ~~~~~{\sf Extension}   &  \multicolumn{2}{c|} {\sf Base}   \\
\hline
 & ~~({\sf Definitions}) & {\sf Terminology} (C_T) & {\sf Constraints} (C_{\Gamma})\\
\hline 
\hline 
0 & & & \begin{array}{l}  
{\underline {\sf organic}} \wedge {\underline {\sf inorganic}} \approx 0\\
{\underline {\sf organic}} \leq {\underline {\sf substance}}  \\ 
{\underline {\sf inorganic}} \leq {\underline {\sf substance}}  \\
\end{array} \\
\hline 
1 & \begin{array}{l} 
{\sf co} \approx {\sf catalyzes}({\sf oxydation}) 
\end{array} 
& \begin{array}{l} 
{\sf cat}\text{-}{\sf oxydation} \approx {\underline {\sf substance}} \wedge {\sf co} \\
\end{array} 
& \begin{array}{l}
{\underline {\sf reaction}} \leq  {\sf oxydation}\\
{\sf cat}\text{-}{\sf oxydation} \leq {\underline {\sf inorganic}}\\
{\sf cat}\text{-}{\sf oxydation} \not\approx 0 \\
\end{array}\\
\hline 
2 & \begin{array}{l} 
{\sf ps} \approx {\sf produces}({\sf substance}) \\
{\sf cr} \approx {\sf catalyzes}({\sf reaction}) \\ 
\end{array}
& \begin{array}{l} {\underline {\sf reaction}} \approx {\sf process} \wedge {\sf ps} \\
{\sf enzyme} \approx {\underline {\sf organic}} \wedge {\sf cr} \\
\end{array} 
& {\sf enzyme} \not\approx 0 \\
\hline 
\end{array}$
}

\medskip
\noindent 
The following instances of the congruence or monotonicity axioms need to be 
considered:  
$${\sf oxydation} \rhd {\sf reaction} \rightarrow {\sf cp} \rhd {\sf cr}, \quad  
\text{ where } \rhd \in \{ \approx, \leq, \geq \}.$$ 
They are not mixed. 
The conjunction of formulae in the base theory is unsatisfiable in the 
theory of semilattices. It can be split into a part $A$ containing only 
concepts in ${\sf AChem}$ and a part $B$ containing only concepts in 
${\sf BioChem}$. 
An interpolant for $A \wedge B$ in the theory of semilattices with 0 
is $I_0 = {\sf substance} \wedge {\sf cr} \leq {\sf inorganic}$. Thus, 
$I = {\sf substance} \wedge 
{\sf catalyzes}({\sf reaction}) \leq {\sf inorganic}$ 
is an interpolant for $A \wedge B$. 
This is an explanation for the inconsistency of $A \wedge B$, and may help 
to find the error more easily than the initial proof of unsatisfiability.
For this we can, for instance, analyze the (shorter) proofs of 
$A \models I$ and 
$B \wedge I \models \perp$ and note that the constraint 
${\sf reaction} \leq {\sf oxydation}$ is used in the proof of 
 $A \models I$.

\subsection{Verification}
\label{appl-verification}
Consider the verification example from 
Section~\ref{motivation-verif}. 
We illustrate our method for generating interpolants for 
a formula corresponding to a path of length 2 from an initial state to an 
unsafe state: 
\begin{eqnarray*} 
G & = &  l < L_{\sf alarm} ~\wedge~ l' \approx {\sf in}(l, t' - t) ~\wedge~ t' \approx k(t) ~\wedge~ l' \geq L_{\sf alarm} ~\wedge~  \\
& & l'' \approx {\sf in}({\sf out}(l', t''_1 - t'), t''_2 - t') ~\wedge~ t''_1 \approx g(t') ~\wedge~ t''_2 \approx h(t')
~\wedge~ \neg l'' < L_{\sf overflow}.
\end{eqnarray*}

\smallskip
{\noindent \em Hierarchic reasoning.} The extension ${\mathcal T}_1$ of 
linear arithmetic with the clauses ${\mathcal K}$ 
in Section~\ref{motivation} is 
local, so to prove $G \models_{{\mathcal T}_1} \perp$ 
it is sufficient to 
consider ground instances ${\mathcal K}[G]$  in which all 
extension terms already  occur in $G$. 
After flattening and purifying ${\mathcal K}[G] \wedge G$, 
we separate the problem into a definition part ({\sf Extension})  
and a base part $G_0 \wedge {\mathcal K}_0$. 
By Theorem~\ref{thm:hierarchic} \cite{Sofronie-cade-05}, 
the problem can be reduced to testing the satisfiability in the base theory of
the conjunction $G_0 \wedge {\mathcal K}_0 \wedge {\sf Con}_0$.
As this conjunction is unsatisfiable with respect to ${\mathcal T}_0$, $G$ is unsatisfiable.
 
\medskip
\noindent $\begin{array}{|l|llll|}
\hline 
 ~~{\sf Extension}   &  \multicolumn{4}{c|} {\sf Base}   \\
\hline
 ({\sf Definitions})  &  G_0    & &  & {\mathcal K}_0 \wedge {\sf Con}_0 \\
\hline 
\hline 
l' \approx {\sf in}(l, e_1) &   l < L_{\sf alarm}  &   & ~{\mathcal K}_0: &  l < L_{\sf alarm} \wedge 0 \leq e_1 \leq \Delta t \rightarrow l' < L_{\sf overflow}\\

c^1_2 \approx {\sf out}(l', e^1_2)  & l' \geq L_{\sf alarm}  &   &  & c^1_2 < L_{\sf alarm} \wedge 0 \leq e^2_2 \leq \Delta t \rightarrow l'' < L_{\sf overflow} \\
l'' \approx {\sf in}(c^1_2, e^2_2)  & e_1 \approx t' - t  &  & &  l' < L_{\sf overflow} \wedge e^1_2 \geq \delta t \rightarrow c^1_2 < L_{\sf alarm} \\
t' \approx k(t) & e^1_2 \approx t''_1 - t'  &  & & 0 \leq \delta t \leq t''_1 - t'
\leq t''_2 - t' \leq \Delta t \\
t''_1 \approx g(t')  & e^2_2 \approx t''_2 - t' &  & & 0 \leq t' - t \leq \Delta t \\
t''_2 \approx h(t') & \neg l'' \leq L_{\sf overflow} & &  ~ {\sf Con}_0: & l \approx c^1_2 \wedge e_1 \approx e^2_2 \rightarrow l' \approx l''   \\
\hline 
\end{array}$

\bigskip
\noindent 
{\em Interpolation.} Let $A$ and $B$ be given by:
\[\eqalign{
  A
={}&l<L_{\sf alarm} \,\wedge\,e_1 \approx t' - t \,\land\, 
l' \approx{\sf in}(l, e_1) \,\land\, t' \approx k(l) \cr
  B 
={}&l' \geq L_{\sf alarm}\,\land\, c^1_2 \approx {\sf out}(l', e^1_2) \,\land\, l'' \approx{\sf in}(c^1_2, e^2_2) \,\land\, e^1_2 \approx t''_1 - t' \,\land\, e^2_2 \approx t''_2 - t'\,\land\cr
& t''_1 \approx g(t') \,\land\, t''_2 \approx h(t') \,\land\, \neg l''
< L_{\sf overflow}\ .
}
\]
The set of constants which occur in $A$ is $\{l, t, e_1, l', t' \}$.
In $B$ occur $\{ l', t', c^1_2, l'', t''_1, t''_2, e^1_2,  e^2_2 \}$. 
The shared constants are $l'$ and $t'$. 
To generate an interpolant for $A \wedge B$, 
we partition the clauses in $A_0 \wedge B_0 \wedge {\mathcal K}_0 \wedge {\sf Con}_0 = A_0 \wedge B_0 \wedge {\mathcal K}^A_0 \wedge {\mathcal K}^B_0 \wedge {\sf Con}_0 $, where: 
\[\eqalign{
  A_0 
={}& l{<}L_{\sf alarm} \wedge e_1 \approx t' - t \cr
  B_0 
={}& l' \geq L_{\sf alarm} \wedge  e^1_2 \approx t''_1 - t' \wedge 
e^2_2 \approx t''_2 - t' \wedge 
\neg l''\leq L_{\sf overflow}\cr
  {\mathcal K}^A_0 
={}&(l < L_{\sf alarm} \wedge 0 \leq e_1 \leq \Delta t \rightarrow l' < L_{\sf overflow}) \,\land\, (0 \leq t' - t \leq \Delta t) \cr
  {\mathcal K}^B_0
={}& (c^1_2 < L_{\sf alarm} \wedge 0 \leq e^2_2 \leq
\Delta t \rightarrow l'' < L_{\sf overflow}) \wedge (l' < L_{\sf
  overflow} \wedge e^1_2 \geq \delta t \rightarrow c^1_2 < L_{\sf
  alarm}) \cr &\wedge (0 \leq \delta t \leq t''_1 - t'\leq t''_2 -
t' \leq \Delta t)\ .
}
\]
The clause in ${\sf Con}_0$ is mixed.  Since already the conjunction
of the formulae in $A_0 \wedge B_0 \wedge {\mathcal K}^A_0 \wedge
{\mathcal K}^B_0$ is unsatisfiable, ${\sf Con}_0$ is not needed to
prove unsatisfiability.  The conjunction of the formulae in $A_0
\wedge B_0 \wedge {\mathcal K}^A_0 \wedge {\mathcal K}^B_0$ is
equivalent to $A'_0 \wedge B'_0$, where
\[\eqalign{
  A'_0
\,=\,{}& l {<} L_{\sf alarm} \wedge e_1 \approx t' - t \wedge (0 \leq t' - t \leq \Delta t) \wedge l'  < L_{\sf overflow} \cr
  B'_0
\,=\,{}& l' > L_{\sf alarm} \wedge e^1_2 \approx t''_1 - t' \wedge 
e^2_2 \approx t''_2 - t' \wedge (0 \leq \delta t \leq t''_1 - t'\leq t''_2 - t' \leq \Delta t) \wedge \cr
& \neg l'' < L_{\sf overflow} \wedge \neg c^1_2 < L_{\sf alarm} \wedge \neg l'< L_{\sf overflow}.
}
\]
The interpolant 
for $A'_0 \wedge B'_0$ is $l' < L_{\sf overflow}$, which 
is also an interpolant for $A \wedge B$. 

\noindent The abstraction defined in Section~\ref{motivation-verif} 
can then be refined by introducing another predicate
$L' < L_{\sf overflow}$.

\section{Conclusions}
\label{conclusions}
\noindent We presented a method for 
obtaining simple interpolants in 
theory extensions. We identified situations in which it is 
possible to do this in a hierarchical 
manner, by using a prover and a procedure for 
generating interpolants in the base theory as ``black-boxes''.
This allows us to use the properties of 
${\mathcal T}_0$ (e.g.\ the form of interpolants) 
to control the form of interpolants in the extension ${\mathcal T}_1$.
We discussed applications of interpolation in verification and knowledge 
representation.

The method we presented can be applied to a class of theories which is more 
general than that considered in 
McMillan \cite{McMillanProver04}  
(extension of linear rational arithmetic 
with uninterpreted function symbols). Our method is orthogonal to 
the method for generating interpolants for combinations of 
theories over disjoint signatures from Nelson-Oppen-style unsatisfiability 
proofs proposed by Yorsh and Musuvathi in 
\cite{Yorsh-Musuvathi-cade-2005}, as it allows us to consider 
combinations of theories over non-disjoint signatures.

The hierarchical interpolation method presented here was in particular 
used for efficiently computing interpolants 
in the special case of the extension of linear arithmetic with 
free function symbols in \cite{RybalchenkoSofronie06};  
the algorithm we used in that paper (on which an implementation is 
based) differs a bit from the one presented here
in being tuned to the constrained based approach used in 
\cite{RybalchenkoSofronie06}.
The implementation was integrated into the predicate discovery
procedure of the software verification tools \blast~\cite{BLASTPOPL04}
and \armc~\cite{ARMC-Saga}.  
First tests suggest that the performance of 
our method is of the same order of magnitude
as the methods which construct interpolants from proofs, 
and considerably faster on many examples. In addition, our method 
can handle systems which pose problems to other
interpolation-based provers: we can handle 
problems containing both strict and nonstrict inequalities, and it allows us to
verify examples that require predicates over up to four variables.
Details about the implementation 
and benchmarks for the special case of linear arithmetic + free function 
symbols are described in \cite{RybalchenkoSofronie06}.

\smallskip
Although the method we presented here is based on a hierarchical reduction 
of proof tasks in a local extension of a given theory ${\mathcal T}_0$ 
to proof tasks in ${\mathcal T}_0$, the results presented in 
Section~\ref{hierarchic} (in particular the separation technique described in 
Proposition~\ref{prop:separation}) and in Section~\ref{procedure} 
also hold for non-purified formulae (i.e.\ they also hold if we do not 
perform the step of introducing new constant names $c_{f(d)}$ for the ground 
terms $f(d)$ which occur in the problem or during the separation process). 
Depending on the properties of ${\mathcal T}_0$, techniques for reasoning 
and interpolant generation in the extension of ${\mathcal T}_0$ with 
free function symbols e.g.\ within state of the art SMT solvers can 
then be used. We can, therefore, use the results in Sections~\ref{hierarchic}
and~\ref{procedure} to extend in a natural way existing methods for 
interpolant computation 
which take advantage of state of the art SMT technology (cf.\ e.g.\ 
\cite{cimatti-tacas08}) to the more complex types of theory extensions 
with sets of axioms of type~(\ref{general-form}) we considered here.  

\smallskip
An immediate application of our method 
is to verification by abstraction-refinement; 
there are other potential applications (e.g.\ 
goal-directed overapproximation for achieving faster termination, or 
automatic invariant generation) which we would like to study. 
We would also like to analyze in more detail the applications to 
reasoning in complex knowledge bases.

\smallskip
\noindent {\bf Acknowledgements.} 
I thank Andrey Rybalchenko for interesting discussions. 
I thank the referees for helpful comments.

\smallskip
\noindent This work was 
  partly supported by the German Research
  Council (DFG) as part of the Trans\-re\-gional
  Collaborative Research Center ``Automatic
  Verification and Analysis of Complex
  Systems'' (SFB/TR 14 AVACS). See
  \texttt{www.avacs.org} for more
  information.

\appendix
\section{Amalgamation and interpolation}
\label{app:amalg-interp}

\noindent There exist results which relate ground interpolation to 
amalgamation or the injection transfer property 
\cite{Jonsson65,Bacsich75,Wronski86} 
and thus allow us to recognize 
many theories with ground interpolation. 

If $\Pi = (\Sigma, {\sf Pred})$ is a signature and 
${\mathcal A}, {\mathcal B}$ are $\Pi$-structures, 
we say that: 
\begin{enumerate}[$\bullet$]
\item a map 
$h : {\mathcal A} \hookrightarrow {\mathcal B}$ is a {\em homomorphism} 
if it preserves the truth of positive literals, i.e.\ has the property 
that if $f_A(a_1, \dots, a_n) = a$ then $f_B(h(a_1), \dots, h(a_n)) = h(a)$, 
and if $P_A(a_1, \dots, a_n)$ is true then $P_B(h(a_1), \dots, h(a_n))$ is true.
\item a map 
$i : {\mathcal A} \hookrightarrow {\mathcal B}$ is an {\em embedding} if 
it preserves the truth of both positive and negative literals, i.e.\ 
$P_A(a_1, \dots, a_n)$ is true (in ${\mathcal A}$) if and only if 
$P_B(i(a_1), \dots,i(a_n))$ is true (in ${\mathcal B}$) for any predicate 
symbol, including equality. Thus, an embedding is an injective homomorphism
which also preserves the truth of negative literals.
\end{enumerate} 

\begin{defi}
Let $\Pi = (\Sigma, {\sf Pred})$ be a signature, and let ${\mathcal M}$ be a class of $\Pi$-structures. 
\begin{enumerate}[(1)]
\item 
We say that ${\mathcal M}$ has the {\em amalgamation property} (AP) if 
for any ${\mathcal A}, {\mathcal B}_1, {\mathcal B}_2 \in {\mathcal M}$  
and any 
embeddings
$i_1 : {\mathcal A} \hookrightarrow {\mathcal B}_1$ and  
$i_2 : {\mathcal A} \hookrightarrow {\mathcal B}_2$ there exists a structure 
${\mathcal C} \in {\mathcal M}$ and 
embeddings 
$j_1 : {\mathcal B}_1 \hookrightarrow {\mathcal C}$ and 
$j_2 : {\mathcal B}_2 \hookrightarrow {\mathcal C}$ such that 
$j_1 \circ i_1 = j_2 \circ i_2$. 
\item
${\mathcal M}$ has the {\em injection transfer property} (ITP) if 
for any ${\mathcal A}, {\mathcal B}_1, {\mathcal B}_2 \in {\mathcal M}$,  
any 
embedding 
$i_1 : {\mathcal A} \hookrightarrow {\mathcal B}_1$ and  
any homomorphism 
$f_2 : {\mathcal A} \rightarrow {\mathcal B}_2$ there exists a structure 
${\mathcal C} \in {\mathcal M}$, a homomorphism 
$f_1 : {\mathcal B}_1 \rightarrow {\mathcal C}$ and an 
embedding 
$j_2 : {\mathcal B}_2 \hookrightarrow {\mathcal C}$ such that 
$f_1 \circ i_1 = j_2 \circ f_2$.
\end{enumerate}
\end{defi}
\begin{defi}
An equational theory ${\mathcal T}$ 
(in signature $\Pi = (\Sigma, {\sf Pred})$ where 
${\sf Pred} = \{ \approx \}$) has 
the {\em equational interpolation property} if
whenever 
$$\bigwedge_i A_i({\overline a}, {\overline c}) \wedge 
\bigwedge_j B_j({\overline c}, {\overline b}) \wedge 
\neg B({\overline c}, {\overline b}) \models_{\mathcal T} \perp,$$ where 
$A_i$, $B_j$ and $B$ are ground atoms,  
there exists a conjunction $I({\overline c})$ of ground atoms
containing only 
the constants ${\overline c}$ occurring both in 
$\bigwedge_i A_i({\overline a}, {\overline c})$ and $\bigwedge_j B_j({\overline c}, {\overline b}) \wedge \neg B({\overline c}, {\overline b})$, 
such that  
$\bigwedge_i A_i({\overline a}, {\overline c})  \models_{\mathcal T}
I({\overline c}) \text{ and } I({\overline c}) \wedge
\bigwedge_j B_j({\overline c}, {\overline b}) \models_{\mathcal T}
B({\overline c}, {\overline b})$
\label{equational-interpolation}
\end{defi}
\begin{thm}[\cite{Jonsson65,Bacsich75,Wronski86}] 
Let ${\mathcal T}$ be a universal theory. 
Then: 
\begin{enumerate}[\em(1)]
\item ${\mathcal T}$ has ground interpolation  
if and only if 
${\sf Mod}({\mathcal T})$ has (AP) \cite{Bacsich75}. 
In addition, we can guarantee that if $\phi$ is positive 
then the interpolant 
of $\phi \wedge \psi$ is positive if and only if    
${\sf Mod}({\mathcal T})$ has the injection transfer property \cite{Bacsich75}.  

\item If ${\mathcal T}$ is an equational theory, then 
${\mathcal T}$ has the equational interpolation property if and only if  
${\sf Mod}({\mathcal T})$ has the injection transfer property \cite{Wronski86}.
\end{enumerate}
\label{amalgamation-interpolation}
\end{thm}

\noindent Theorem~\ref{amalgamation-interpolation} can be used to prove that 
many equational theories have  ground interpolation: 

\begin{thm} 
The following theories allow ground interpolation\footnote{In fact, 
the theories (1) and (4) allow  equational interpolation. Similar results were also established for (2)  in \cite{RybalchenkoSofronie06}.}: 
\begin{enumerate}[\em(1)]
\item The theory of pure equality (without function symbols). 
\item Linear rational and real arithmetic.
\item The theory of posets.
\item The theories of (a) Boolean algebras, (b) semilattices, (c) 
distributive lattices.
\end{enumerate}
\label{app:ground-interp-eq-th}
\end{thm}
\Proof (1), (2), (3) are well-known (for (2) we refer for instance to 
\cite{McMillanProver04} or \cite{Yorsh-Musuvathi-cade-2005}). 
For proving (4) we use the fact that 
if a universal theory has a positive algebraic completion 
then it has the injection transfer property \cite{BaaderGhilardi-05}. 
All theories in (4) are equational theories; 
by results in \cite{Wronski86}, for equational theories the injection transfer property is equivalent to 
the equational interpolation property. 
With these remarks, (4)(a) follows from the fact that 
any Gaussian theory is its own positive algebraic completion
\cite{BaaderGhilardi-05}, and (4)(b),(c) from the fact that 
the theory of semilattices and that of distributive 
lattices have  a positive algebraic completion \cite{BaaderGhilardi-05}.
\qed

\smallskip
\noindent Similarly it can be proved that 
the equational classes of 
(abelian) groups and lattices have ground interpolation.

\section{Proof of Theorem~\ref{examples-local}}
\label{app:proof-ex-local-ext}

\noindent 
{\bf Theorem~\ref{examples-local}}
{\em 
We consider the following base theories ${\mathcal T}_0$: 
\begin{enumerate}[\em(1)]
\item ${\mathcal P}$ (posets), 
\item ${\mathcal TO}$ (totally-ordered sets), 
\item ${\sf SLat}$ (semilattices), 
\item ${\sf DLat}$ (distributive lattices), 
\item ${\sf Bool}$ (Boolean algebras). 
\item the theory ${\mathbb R}$ of reals resp.\ 
${\sf LI}(\mathbb R)$ (linear arithmetic over ${\mathbb R}$), or 
the theory ${\mathbb Q}$ of rationals resp.\  
${\sf LI}(\mathbb Q)$ (linear arithmetic over ${\mathbb Q}$),   
or (a subtheory of) the theory of integers (e.g.\ Presburger
arithmetic). 
\end{enumerate}
The following theory extensions are local:
\begin{enumerate}[\em(a)] 
\item Extensions of any theory ${\mathcal T}_0$ 
for which $\leq$ is reflexive with functions satisfying boundedness 
$({\sf Bound}^t(f))$ or guarded boundedness $({\sf GBound}^t(f))$ conditions 

\smallskip
$({\sf Bound}^t(f)) \quad  \quad \forall x_1, \dots, x_n (f(x_1, \dots, x_n) \leq t(x_1, \dots, x_n))$ 

$({\sf GBound}^t(f)) \quad \forall x_1, \dots, x_n (\phi(x_1, \dots, x_n) \rightarrow f(x_1, \dots, x_n) \leq t(x_1, \dots, x_n)),$

\smallskip
\noindent 
where $t(x_1, \dots, x_n)$ is a term in the base signature $\Pi_0$ and 
$\phi(x_1, \dots, x_n)$ a conjunction of literals in the signature $\Pi_0$, 
whose variables are in $\{ x_1, \dots, x_n \}$.

\smallskip 
\item Extensions of any theory  ${\mathcal T}_0$  in (1)--(6)  
with ${\sf Mon}(f) \wedge {\sf Bound}^t(f)$, if $t(x_1, \dots, x_n)$ is 
a term in the base signature $\Pi_0$ in the variables $x_1, \dots, x_n$ 
such that for every model of ${\mathcal T}_0$ the associated function is
monotone in the variables $x_1, \dots, x_n$.

\smallskip
\item Extensions of any theory in  (1)--(6) 
with functions satisfying ${\sf Leq}(f,g) \wedge {\sf Mon}(f)$.

\smallskip
$({\sf Leq}(f,g)) \quad \forall x_1, \dots, x_n (\bigwedge_{i = 1}^n x_i \leq y_i \rightarrow f(x_1, \dots, x_n) \leq g(y_1, \dots, y_n))$

\smallskip
\item Extensions of any totally-ordered theory above (i.e.\ (2) and (6))
with functions satisfying 
${\sf SGc}(f,g_1, \dots, g_n) \wedge {\sf Mon}(f, g_1, \dots, g_n)$.

\smallskip
$({\sf SGc}(f,g_1, \dots, g_n)) \quad \forall x_1,\dots, x_n, x ( \bigwedge_{i = 1}^n x_i  \leq g_i(x) \rightarrow f(x_1, \dots, x_n) \leq x)$

\smallskip
\item Extensions of  any theory in (1)--(3) 
with functions satisfying ${\sf SGc}(f,g_1) \wedge {\sf Mon}(f, g_1)$.
\end{enumerate}

\smallskip
\noindent 
All the extensions above satisfy condition $({\sf Loc^f})$.
} 

\smallskip
\Proof In what follows we will denote by $\Pi_0$ the signature of the base 
theory ${\mathcal T}_0$, and with $\Sigma_1$ the extension functions, namely 
$f$ for cases (a) and (b), $f, g$ for case (c), $f, g_1, \dots, g_n$ for 
case (d) and $f, g_1$ for case (e). 

\medskip
\noindent (a)  Let $(P, f_P)$ be a partial $\Pi$-structure which  weakly 
satisfies ${\sf Bound}^t(f)$, such that $P \in {\sf Mod}({\mathcal T}_0)$ and 
$f_P : P^n \rightarrow P$ is partial. 
Let $A = (P, f_A)$ be a total $\Pi$-structure 
with the same support as $P$, where: 
$$f_A(x_1, \dots, x_n) = \left \{ \begin{array}{ll} 
f_P(x_1, \dots, x_n) & \text{ if } f_P(x_1, \dots, x_n) \text{ defined} \\
t(x_1, \dots, x_n) & \text{ otherwise}. 
\end{array}
\right.$$
Then $A$ satisfies ${\sf Bound}^t(f)$. 
Let $i : (P, f_P) \rightarrow (A, f_A)$ be the identity.
Obviously, $i$ is a $\Pi_0$-isomorphism; and if 
$f_P(x_1, \dots, x_n)$ is defined then 
$i(f_P(x_1, \dots, x_n)) = f_P(x_1, \dots, x_n) = f_A(x_1, \dots, x_n)$.
Similar arguments also apply to ${\sf GBound}^t(f)$. 

\medskip
\noindent 
(b) Let $(P, f_P)$ be a partial $\Pi$-structure which  weakly satisfies 
${\sf Bound}^t(f) \wedge {\sf Mon}$, 
such that $P \in {\sf Mod}({\mathcal T}_0)$ and 
$f_P : P^n \rightarrow P$ is partial.  In cases (1)--(3) 
let $A = (\mathcal{OI}(P), {\overline f})$, where 
$\mathcal{OI}(P)$ is the family
 of all order ideals of $P$, and 
$$f_A(U_1, \dots, U_n) = 
{\downarrow}\{ f_P(u_1, \dots, u_i) \mid u_i \in U_i, f_P(u_1, \dots, u_n) 
\text{ defined} \}.$$

\noindent 
$f_A$ is clearly monotone. 
Let $z \in f_A(U_1, \dots, U_n)$. 
Then $z \leq f_P(u_1, \dots, u_n)$
for some $u_i \in U_i$ with  $f_P(u_1, \dots, u_n)$ defined. 
As $P \models_w {\sf Bound}^t(f)$, 
$f_P(u_1, \dots, u_n) \leq t(u_1, \dots, u_n)$. Therefore 
$z \in t(U_1, \dots, U_n)$. The map 
$i : (P, f_P) \rightarrow (A, f_A)$ defined by $i(p) = \downarrow p$ 
is a weak embedding. 

\medskip
Since ${\sf DLat}$ and ${\sf Bool}$ are locally finite, 
results in \cite{Sofronie-cade-05}  show that 
in (4) and (5) it is sufficient 
to assume that $P$ is finite. 
Let $A = (P, f_A)$, where 
$$f_A(x_1, \dots, x_n) = 
\bigvee \{ f_P(u_1, \dots, u_n) \mid u_i \leq x_i, f_P(u_1, \dots, u_n) 
\text{ defined} \}.$$ 
$f_A$ is clearly monotone. 
We prove that it also satisfies the boundedness condition, i.e.\ that 
for all $x_1, \dots, x_n$, $f_A(x_1, \dots x_n) \leq t(x_1, \dots, x_n)$.
By definition, $f_A(x_1, \dots,x_n) = \bigvee \{ f_P(u_1, \dots, u_n) \mid 
u_i \leq x_i, f_P(u_1, \dots, u_n) \text{ defined} \}.$
As $P \models_w {\sf Bound}^t(f)$ and $t$ is monotone, we know that 
$f_P(u_1, \dots, u_n) \leq t(u_1, \dots, u_n) \leq t(x_1, \dots, x_n)$ for all
$u_i \leq x_i$ with  $f_P(u_1, \dots, u_n)$ defined.   
Therefore, 
$$f_A(x_1, \dots, x_n) = 
\bigvee \{ f_P(u_1, \dots, u_n) \mid u_i \leq x_i, f_P(u_1, \dots, u_n) 
\text{ defined} \} \leq  t(x_1, \dots, x_n).$$ 
That the identity $i$ is a weak embedding 
can be proved as before. 

\medskip
\noindent (c) The proof is very similar to the proof of (b). 
We first discuss the case (1)--(3). 
Let $(P, f_P, g_P)$ be a weak partial model of ${\mathcal T}_1$. 
Let $A = (\mathcal{OI}(P), f_A, g_A)$, where $f_A$ is defined as in (b).  
We define $g(U_1, \dots, U_n)$ by
$$g_A(U_1, \dots, U_n) = \left \{ \begin{array}{ll} 
\downarrow g_P(x_1, \dots, x_n) & \text{ if } U_i = \downarrow x_i \text{ and  } g_P(x_1, \dots, x_n) \text{ defined} \\
f_A(U_1, \dots, U_n) & \text{ otherwise}. 
\end{array}
\right.$$
Assume that $U_1 \subseteq V_1, \dots, U_n \subseteq V_n$, and  
let $z \in f_A(U_1, \dots, U_n)$. Then  $z \leq f_P(u_1, \dots, u_n)$
for some $u_i \in U_i \subseteq V_i$ with  
$f_P(u_1, \dots, u_n)$ defined. 
If 
$V_i = \downarrow x_i \text{ and  } g_P(x_1, \dots, x_n) \text{ defined}$, 
then $u_i \leq x_i$ so,  
as $P \models_w {\sf Leq}(f, g)$, we know that 
$f_P(u_1, \dots, u_n) \leq g_P(x_1, \dots, x_n)$. It therefore follows that
in this case $z \in {\downarrow} g_P(x_1, \dots, x_n) = g_A(V_1, \dots, V_n)$. 
Otherwise, $g_A(V_1, \dots, V_n) = f_A(V_1, \dots, V_n)$, 
hence $f_A(U_1, \dots, U_n) \subseteq g_A(V_1, \dots, V_n)$. 

\smallskip
For the cases (4) and (5) we again use the criterion in 
\cite{Sofronie-cade-05}  and Theorem~\ref{rel-loc-embedding}. 
Let $(P, f_P, g_P)$ be a weak partial model of ${\mathcal T}_1$. 
Let $a_0 \in P$ be such that $a_0 \geq f_P(p_1, \dots, p_n)$ whenever 
$ f_P(p_1, \dots, p_n)$ is defined. We define $A = (P, f_A, g_A)$ as follows:
\[\eqalign{
g_A(x_1, \dots, x_n) &= 
\begin{cases}
\downarrow g_P(x_1, \dots, x_n) &\text{if $g_P(x_1, \dots, x_n)$  defined} \cr
a_0 &\text{otherwise}
\end{cases}\cr
f_A(x_1, \dots, x_n)&= \bigvee \{ f_P(u_1, \dots, u_n) \mid u_i \leq
x_i, f_P(u_1, \dots, u_n) \text{ defined} \}\ .
}
\]
$f$ is obviously monotone. In order to prove that the second condition holds, 
we analyze two cases. Assume first that $g_P(y_1, \dots, y_n)$ is undefined.
Then $g_A(y_1, \dots, y_n) = a_0 \geq f_P(u_1, \dots, u_n)$ for all 
$u_i \leq x_i$ with $f_P(u_1, \dots, u_n)$ defined, thus, 
$g_A(y_1, \dots, y_n) = a_0 \geq \bigvee \{ f_P(u_1, \dots, u_n) \mid u_i \leq x_i, f_P(u_1, \dots, u_n) \text{ defined } \} = f_A(x_1, \dots, x_n)$. 
If $g_P(y_1, \dots, y_n)$ is defined, then 
for all 
$u_i \leq x_i$ with $f_P(u_1, \dots, u_n)$ we also have 
$u_i \leq y_i$, so $f_P(u_1, \dots, u_n) \leq g_P(y_1, \dots, y_n) = g_A(y_1, \dots, y_n)$. 
Again, it follows that $g_A(y_1, \dots, y_n) \leq f_A(x_1, \dots, x_n)$. 

\medskip
\noindent 
(d) Let ${\mathcal T}_0$ be the theory of totally ordered sets. 
Assume that $(P, f_P, (g^i_P))$ 
is a totally ordered weak partial model of 
${\sf SGc}(f,g_1, \dots, g_n) \wedge {\sf Mon}(f, g_1, \dots, g_n)$.
Let $A = (\mathcal{OI}(X), f_A, (g^i_A))$, where $f_A$ and $g^i_A$ are extensions of $f_P, g^i_P$ defined as in the proof of (b).
$f_A, g^1_A, \dots, g^n_A$ are obviously monotone. We prove that the 
condition ${\sf SGc}(f,g_1, \dots, g_n)$ holds in $A$.
Assume that $U_i \subseteq g^i_A(V)$ for $i = 1, \dots, n$, and 
let $x \in f_A(U_1, \dots, U_n)$. Then there exist 
$u_i \in U_i = g^i_A(V_i)$ 
such that $f(u_1, \dots, u_n)$ is defined and 
$x \leq f(u_1, \dots, u_n)$. 
As $u_i \in g^i_A(V)$, there exist $v_i \in V$ such that 
$g^i_P(v_i)$ is defined and $u_i \leq g^i_P(v_i)$.
Let $v = {\sf max}(v_1, \dots, v_n)$. 
Then $u_i \leq g^i_P(v)$.  Hence $f_P(u_1, \dots, u_n) \leq v \in V$. 
Therefore, $x \leq f_P(u_1, \dots, u_n) \in V$ so $x \in V$. 
Let $i : P \rightarrow A$ defined by $i(p) = \downarrow p$. 
To show that it is a weak embedding we only have to show that 
if $g_P(x_1, \dots, x_n)$ is defined then 
$i(g_P(x_1, \dots, x_n)) = {\downarrow} g_P(x_1, \dots, x_n) = 
g_A({\downarrow} x_1, \dots, {\downarrow} x_n)$. This is true by the definition of 
$g_A$. 

\medskip
\noindent 
(e) Assume that ${\mathcal T}_0$ is the theory of semilattices. 
The construction in (d) can be applied to this case without problems. 
The proof  
is similar to that of (d) with the difference that if $n = 1$ 
we only have one element $v_1$ so we do not need to compute a maximum
(which for $n \geq 2$ may not exist if the order is not total).

\medskip
The proof of the fact that the remaining theories satisfy $({\sf Loc^f})$ 
is based on the criterion of finite locality given in 
Theorem~\ref{rel-loc-embedding}. The constructions and the proofs are 
similar to those in the proof of (b) resp.\ (c) for the cases (4) and (5). 
Due to the fact that we 
assumed that the definition domain of the extension functions is finite 
$\bigvee \{ f_P(u_1, \dots, u_n) \mid u_i \leq x_i,  f_P(u_1, \dots, u_n) 
\text{ defined} \}$ is a finite join, and thus exists (if $f$ is nowhere 
defined it is sufficient to define it as being everywhere equal to $t$ in 
case (b) or to $g_A$ in case (c)). The fact that the definition domains 
are finite also ensures that in the proof of (c) an element $a_0$ 
(chosen in the definition of $g_A$) with the desired properties always exists.
\qed

\section{Proof of Theorem~\ref{example-assumptions-1-3}}
\label{app:assumptions-examples}

\noindent {\bf Theorem~\ref{example-assumptions-1-3}.}
{\em 
The following theories have ground interpolation and are convex and  
$P$-interpolating with respect to the indicated set $P$ of predicate symbols:
\begin{enumerate}[\em(1)]
\item The theory of ${\mathcal EQ}$ of pure equality without function symbols 
(for $P = \{ \approx \}$). 
\item The theory ${\sf PoSet}$ of posets (for $P = \{ \approx, \leq \}$).
\item Linear rational arithmetic ${\sf LI}({\mathbb Q})$ and 
linear real arithmetic ${\sf LI}({\mathbb R})$ (convex with respect to  $P = \{ \approx \}$, 
strongly $P$-interpolating for $P = \{ \leq \}$). 
\item  The theories ${\sf Bool}$ of Boolean algebras, 
${\sf SLat}$ of semilattices and ${\sf DLat}$ of distributive lattices  
(strongly $P$-interpolating for $P = \{ \approx, \leq \}$).
\end{enumerate}
}

\smallskip
\Proof Note first that if a partially-ordered 
theory is interpolating for $\leq$ it is also for 
$\approx$. Assume that $A \wedge B \models_{\mathcal T} a \approx b$. 
Then $A \wedge B \models_{\mathcal T} a \leq b$ and 
$A \wedge B \models_{\mathcal T} b \leq a$, hence there exist terms $t_1$, $t_2$ 
containing only common constants of $A$ and $B$ such that 
$A \wedge B \models a \leq t_1 \wedge t_1 \leq b$ and 
$A \wedge B \models b \leq t_2 \wedge t_2 \leq a$. 
It follows that $A \wedge B \models t_1 \approx t_2$,  
$A \wedge B \models a \approx t_1 \wedge t_1 \approx b$.   

\smallskip
\noindent (1) and (2): convexity is obvious; the property of being 
$P$-interpolating can be proved by induction on the structure of proofs. 
(3) is known 
(cf.\ e.g.\ \cite{Yorsh-Musuvathi-cade-2005}). A method for computing 
interpolating terms for ${\sf LI}({\mathbb R})$ and ${\sf LI}({\mathbb Q})$
is presented in \cite{RybalchenkoSofronie06}.

\smallskip
\noindent 
(4) This is a constructive proof based on ideas from 
\cite{Sofronie-cade-1999,Sofronie-jsc-2003}. The results presented there 
show, as an easy particular case, that one can reduce the problem of checking 
the satisfiability 
of a conjunction $\Gamma$ of unit clauses with respect to one of the theories above 
to checking the satisfiability of a conjunction ${\sf Ren}_{\Gamma} \wedge {\sf P}_{\Gamma} \wedge {\sf N}_{\Gamma}$ obtained by introducing a propositional 
variable $P_e$ for each subterm $e$ occurring in $\Gamma$, 
a set of renaming rules of the form 
\begin{eqnarray*}
P_{e_1 ~{\sf op}~ e_2} & \!\!\!\leftrightarrow   P_{e_1} ~{\sf op}~ P_{e_2} ~~~~~~~  & {\sf op} \text{ binary Boolean operation} \\
P_{\neg e} & \!\!\!\leftrightarrow  \neg P_e~~~~~~~~~~~~~~~  &  \text{ in the case of } {\sf Bool}, 
\end{eqnarray*} and translations of the positive resp.\ negative part of 
$\Gamma$:
\begin{eqnarray*}
P_{s} & \!\!\!\leftrightarrow  P_{s'}~~~~~~~ &  s \approx s' \in \Gamma \\
\neg (P_{s} & \!\!\!\leftrightarrow  P_{s'})~~~~~  &  s \not\approx s' \in \Gamma.
\end{eqnarray*}

\medskip
\noindent 
(a) The convexity of the theory of Boolean algebras with respect to 
$\approx$ 
follows from the fact that this is an equational class; convexity with respect 
to $\leq$ follows from the fact that $x \leq y$ if and only if $x \wedge y \approx x$.
We prove that the theory of Boolean algebras is $\leq$-interpolating, 
i.e. that if $A$ and $B$ are two conjunctions of literals 
and $A \wedge B \models_{\sf Bool} a \leq b$, where $a$ is a constant occurring in $A$ 
and not in $B$ and $b$ a constant occurring in $B$ and not in $A$, then 
there exists a term containing only common constants in $A$ and $B$ such that 
$A \models_{\sf Bool} a \leq t$ and $B \models_{\sf Bool} t \leq b$. 
We can assume without loss of generality that $A$ and $B$ consist only
of atoms (otherwise one moves the negative literals to the right and
uses convexity).  $A \wedge B \models_{\sf Bool} a \leq b$ if and only
if the following conjunction of literals in propositional logic is
unsatisfiable:

$$\begin{array}{ccc} 
\begin{array}{lrcl}
{\sf (Ren(\wedge))}\hspace{.3cm} & P_{e_1 \wedge e_2} & \leftrightarrow & P_{e_1} \wedge P_{e_2} \\
{\sf (Ren(\vee))} & P_{e_1 \vee e_2} & \leftrightarrow & P_{e_1} \vee P_{e_2} \\ 
{\sf (Ren(\neg))} & P_{\neg e} & \leftrightarrow & \neg P_{e}  \\ 
{\sf (P)} & P_{e_1} & \leftrightarrow & P_{e_2} ~~~ e_1 \approx e_2 \in A\\
{\sf (N)} & & P_a & \\\\
&&&\quad\hbox to 0 pt{\hss for all $e$, $e_1$, $e_2$ subterms in $A$\hss} 
\end{array}
&& 
\begin{array}{rcl}
 P_{g_1 \wedge g_2} & \leftrightarrow & P_{g_1} \wedge P_{g_2} \\
P_{g_1 \vee g_2} & \leftrightarrow & P_{g_1} \vee P_{g_2} \\
P_{\neg g} & \leftrightarrow & \neg P_{g} \\
  P_{g_1} & \leftrightarrow & P_{g_2} ~~~ g_1 \approx g_2 \in B\\
& \neg P_b & \\\\
&&\ \hbox to 0 pt{\hss for all $g$, $g_1$, $g_2$ subterms in $B$\hss}
\end{array}\\
\end{array}$$

\smallskip
\noindent 
We obtain an unsatisfiable set of clauses 
$(N_A \wedge P_a) \wedge (N_B \wedge \neg P_b) \models \perp$.
Propositional logic allows interpolation, so there exists an interpolant 
$I = f(P_{e_1}, \dots, P_{e_n})$, which is a Boolean combination (say in CNF) 
of the common propositional variables occurring in 
$N_A$ and $N_B$ such that 
$$(N_A \wedge P_a) \models I \quad \text{ and } \quad 
(N_B \wedge \neg P_b) \wedge I \models \perp.$$
\noindent 
But then $A \models_{\sf Bool} a \leq f(e_1, \dots, e_n)$ and 
$B \models_{\sf Bool} f(e_1, \dots, e_n) \leq b$. 

\medskip 
\noindent 
(4)(b) The proof is similar to that of (4)(a) with the difference that in the 
renaming rules in the structure-preserving translation to clause form 
only the conjunction rules apply, hence $N_A$ and $N_B$ are sets of 
non-negative Horn clauses. 
We can saturate $N_A \cup P_a$ under resolution with 
selection on the negative literals in linear time.
The saturated set $N_A^*$ of clauses contains 
all unit clauses $P_e$ where $e$ is subterm of $A$ with 
$A \models_{\sf SLat} a \leq e$. 
Only  unit positive clauses $P_e$ where $e$ occurs in both $A$ and $B$ 
can enter into resolution inferences with clauses in $N_B \cup \neg P_b$
and lead to a contradiction. Thus we proved that 
\begin{eqnarray*}
{\displaystyle \bigwedge} \{ P_e \mid 
A \models_{\sf SLat} a \leq e, e \text{ common subterm} \} 
\wedge N_B \wedge \neg P_b  \models \perp. \label{clauses}
\end{eqnarray*}
\noindent 
This is equivalent to 
$B \models_{\sf SLat} t  \leq b$, where 
\[t = \bigwedge \{ e \mid A \models_{\sf SLat} a \leq e, e \text{
common } \text{ subterm of } A\text{ and } B\}\ .
\] 
Obviously, $A \models_{\sf SLat} a \leq t$.

\medskip
\noindent 
(4)(c) The case of distributive lattices can be treated similarly. 
Due to the fact that in this case the renaming rules 
for $\vee$ and $\wedge$ are taken into account, the sets $N_A$ and $N_B$ 
are not Horn. We adopt the same negative selection strategy. 
When saturating $N_A \cup P_a$ a finite set of positive clauses is 
generated, namely of the form $P_{e_1} \vee \dots \vee P_{e_n}$ where 
$A \models_{\sf DL} a \leq (e_1 \vee \dots \vee e_n)$. 
We consider a total ordering on the propositional variables 
where $P_e$ is larger than $P_g$ if $e$ occurs in $A$ and not in $B$ 
and $g$ occurs in both $A$ and in $B$. Then the only inferences which can lead 
to a contradiction with $N_B \cup \neg P_b$ are those between the clauses 
in $N_A^*$ which only contain common propositional variables. 
Thus we proved that 
\begin{eqnarray*}
{\displaystyle \bigwedge} \{ \bigvee P_{e_i} \mid 
A \models_{\sf DL} a \leq \bigvee e_i, e_i \text{ common terms} \} 
\wedge N_B \wedge \neg P_b \models \perp. \label{distr-clauses}
\end{eqnarray*}
\noindent This is equivalent to 
$B \models_{\sf DL} t \leq b$, where 
$t = \bigwedge \{ \bigvee e_i  \mid A \models_{\sf DL} a \leq \bigvee e_i, 
\text{ ~where~ all~ } 
e_i \text{ are} \\
\text{common subterms of } A \text{ and } B \}$. Obviously, $A \models_{\sf DL} a \leq t$.  
\qed

\end{document}